\documentclass[11pt]{article}

\usepackage[T1]{fontenc}
\usepackage{lmodern}

\usepackage{amsmath}
\usepackage{amsfonts}

\usepackage[mathscr]{euscript}

\usepackage[cm]{fullpage}

\usepackage{xcolor}
\definecolor{plum}{rgb}{0.36078, 0.20784, 0.4}
\definecolor{chameleon}{rgb}{0.30588, 0.60392, 0.023529}
\definecolor{cornflower}{rgb}{0.12549, 0.29020, 0.52941}
\definecolor{scarlet}{rgb}{0.8, 0, 0}
\definecolor{brick}{rgb}{0.64314, 0, 0}	

\usepackage[		%
	pdftitle={Holographic Renormalization for Asymptotically Lifshitz Spacetimes},	%
	pdfauthor={Robert Mann, Robert McNees},	%
	colorlinks=true,	%
	linkcolor=cornflower,	%
	citecolor=brick,	%
	urlcolor=brick	%
	]{hyperref}

\usepackage{tensor}

\numberwithin{equation}{section}

\newcommand{\ts}[1]{\textrm{\tiny #1}}
\newcommand{\ms}[1]{\textrm{\tiny $#1$}}

\newcommand{\LO}{\textrm{\tiny $(0)$}}
\newcommand{\NLO}{\textrm{\tiny $(2)$}}
\newcommand{\NNLO}{\textrm{\tiny $(4)$}}

\newcommand{\email}[1]{\href{mailto:#1}{\tt \textcolor{cornflower}{#1}}}

\newcommand\nts{\negthickspace}
\newcommand\bns{\nts \nts}

\newcommand\defeq{\mathrel{\mathop:}=}

\newcommand\RR{\mathcal{R}}
\newcommand\OO{\mathcal{O}}
\newcommand\MM{\mathcal{M}}

\newcommand\dM{\partial\mathcal{M}}
\newcommand\FF{F}
\newcommand\DD{\mathscr{D}}

\newcommand\CC{\mathcal{C}}
\newcommand\EE{\mathcal{E}}
\newcommand\eps{\epsilon}
\newcommand\dd{\textrm{d}}

\newcommand\Lie{\pounds} 

\newcommand\lsq{\ell^{\,2}}

\renewcommand\AA{\mathcal{A}}

\newcommand\RS{\tensor[^4]{R}{}}

\newcommand{\CD}{\tensor[^4]{\nabla}{}}

\begin{document}


\begin{titlepage}

\begin{flushright}
\today 
\end{flushright}
~\vspace{2cm}

\begin{center}
	{\bf \Large Holographic Renormalization for Asymptotically Lifshitz Spacetimes}
\end{center}

\vspace{.5cm}

\begin{center}
Robert B. Mann $^{a, b}$ and Robert McNees $^{c}$

\vspace{.5cm}{\small {\textit{$^{a}$Dept. of Physics, University of
Waterloo, Waterloo, Ontario N2L 3G1, Canada}}}\\
\vspace{2mm} {\small {\textit{$^{b}$Perimeter Institute for Theoretical Physics, 31 Caroline Street North, Waterloo,
Ontario N2L 2Y5, Canada}}}\\
\vspace{2mm} {\small {\textit{$^{c}$Loyola University Chicago, Department of Physics, Chicago, IL 60660}}}\\
\vspace*{0.5cm}
\email{rbmann@sciborg.uwaterloo.ca}\,,
\email{rmcnees@luc.edu}
\end{center}
\vspace{1cm}

\begin{abstract}
A variational formulation is given for a theory of gravity coupled to a massive vector in four dimensions, with Asymptotically Lifshitz boundary conditions on the fields. For theories with critical exponent $z=2$ we obtain a well-defined variational principle by explicitly constructing two actions with local boundary counterterms. As part of our analysis we obtain solutions of these theories on a neighborhood of spatial infinity, study the asymptotic symmetries, and consider different definitions of the boundary stress tensor and associated charges. A constraint on the boundary data for the fields figures prominently in one of our formulations, and in that case the only suitable definition of the boundary stress tensor is due to Hollands, Ishibashi, and Marolf. Their definition naturally emerges from our requirement of finiteness of the action under Hamilton-Jacobi variations of the fields. A second, more general variational principle also allows the Brown-York definition of a boundary stress tensor.

\end{abstract}

\end{titlepage}

\tableofcontents

~

\section{Introduction}
\label{sec:Intro}

An interesting extension of AdS/CFT is the study of models that arise in condensed matter physics, particularly  those describing quantum critical systems \cite{Hartnoll:2009sz, Rokhsar:1988zz, Ardonne:2003wa, Vishwanath}.  Such models, which are difficult to study using traditional methods based on weakly interacting quasiparticles and broken symmetry, may exhibit non-relativistic scaling symmetries of the form
\begin{gather}\label{LifshitzScaling}
	t \to \lambda^{z} \,t \qquad \qquad \vec{x} \to \lambda \, \vec{x}
\end{gather}
with dynamical critical exponent $z \neq 1$. These symmetries, present at quantum critical points, provide a  strong kinematic connection to some versions of the AdS/CFT correspondence. The transformations \eqref{LifshitzScaling}, known as Lifshitz scaling, are encoded in the asymptotic symmetry group of the conjectured gravitational dual theory.

An early manifestation of this idea  is based on four-dimensional gravity with a vector and a 2-form that interact via a topological coupling \cite{Kachru:2008yh}. However an equivalent (and simpler, for our purposes) formulation is obtained by integrating out the 2-form \cite{Copsey:2010ya}. The result is gravity coupled to a massive vector, described by the Lagrangian
	\begin{gather}\label{VectorFormulationLagrangian}
		L = \frac{1}{2\kappa^2} \, \sqrt{-g} \left( \RS - 2\,\Lambda - \frac{1}{4}\,F^{\mu\nu}\,F_{\mu\nu} - \frac{m^2}{2}\,A^{\mu} A_{\mu}\right) ~.
\end{gather} 
This Lagrangian yields the following equations of motion for the metric and vector
\begin{gather}\label{Einstein}
	\RS_{\mu\nu} - \frac{1}{2}\,g_{\mu\nu}\,\RS = -\Lambda\,g_{\mu\nu} + \frac{1}{2}\,\left(F_{\mu}{}^{\lambda}\,F_{\nu\lambda} - \frac{1}{4}\,g_{\mu\nu}\,F^{\lambda\kappa} F_{\lambda\kappa} \right) + \frac{m^2}{2}\,\left( A_\mu A_\nu - \frac{1}{2}\,g_{\mu\nu}\,A^{\lambda} A_{\lambda}\right) \\ \label{Proca}
	\CD^{\mu} F_{\mu\nu} = m^2 \, A_{\nu} ~.
\end{gather}
With the appropriate boundary conditions, the equations of motion admit solutions that realize the non-relativistic scaling symmetries. The simplest solution of this kind is 
\begin{align}\label{SimpleLifshitzMetric}
		ds^{2} = &\,\, \left(\frac{\ell}{r}\right)^{2} dr^{2} - \left(\frac{r}{\ell}\right)^{2z} dt^{2} + \left(\frac{r}{\ell}\right)^{2} d\vec{x}^{\,2} \\ \label{SimpleLifshitzVector}
		A_{\mu} dx^{\mu} = & \,\, \left(\frac{r}{\ell}\right)^{z} \, \sqrt{\frac{2(z-1)}{z}} \, dt ~.
	\end{align}	
The critical exponent $z$ and the length scale $\ell$ that characterize the solution are related to the parameters of the theory by
	\begin{gather}
		m = \frac{\sqrt{2\,z}}{\ell} \quad \quad \quad \Lambda = - \frac{z^2 + z + 4}{2\,\ell^{2}}
	\end{gather}
For $z=1$, the vector field vanishes and the metric \eqref{SimpleLifshitzMetric} describes the Poincar\'e  patch of AdS$_4$. Other solutions with the same boundary conditions describe localized excitations of a massive vector (i.e., excitations with compact support) on an asymptotically AdS$_4$ spacetime. The asymptotic symmetry group of the properly formulated theory is the three-dimensional conformal group SO(3,2). However, when $z\neq 1$ the symmetries change. The asymptotic symmetries of these `Lifshitz' solutions \cite{Mann:2009yx,Lsolutions} include the non-relativistic scaling transformations \eqref{LifshitzScaling} (noted previously in other contexts \cite{Koroteev:2007yp}), translations, and spatial rotations, but not boosts. The case $z=2$ is of particular interest, because certain models of strongly correlated electrons are invariant under the transformations $t \to \lambda^{2}\,t$, $\vec{x} \to \lambda\,\vec{x}$.

As in AdS/CFT, the on-shell action is an important tool in studying the properties of the dual Condensed Matter models. However, simply integrating the Lagrangian \eqref{VectorFormulationLagrangian} over spacetime and adding the Gibbons-Hawking-York term \cite{York:1972sj,Gibbons:1976ue} at spatial infinity
\begin{align}\label{VectorFormulation}
	I = & \,\, \frac{1}{2\kappa^2} \int_{\MM} \nts \nts \textrm{d}^{4}x \, \sqrt{-g} \left( \RS - 2\,\Lambda - \frac{1}{4}\,F^{\mu\nu}\,F_{\mu\nu} - \frac{m^2}{2}\,A^{\mu} A_{\mu}\right) 
	+ \frac{1}{\kappa^2} \int_{\dM} \bns \textrm{d}^{3}x\,\sqrt{-h}\,K
\end{align}
does not give an appropriate action. An immediate problem with this action is that it exhibits divergences and other undesirable properties when evaluated on solutions asymptotic to \eqref{SimpleLifshitzMetric}-\eqref{SimpleLifshitzVector}. This is because -- as is often the case when one makes the jump from \eqref{VectorFormulationLagrangian} to \eqref{VectorFormulation} -- the EOM are not actually stationary points of the proposed action. The variation of this action contains surface terms that do not vanish unless the field variations satisfy more restrictive fall-off conditions than the solutions themselves.

This is a common problem when formulating gravitational theories on non-compact spacetimes, where a clearly defined boundary value problem is not automatically equivalent to a well-defined variational principle. For example, a variational principle for asymptotically flat gravity requires a new boundary term in addition to the usual Gibbons-Hawking-York term. Otherwise the action is not stationary under generic $1/r$ deformations of the Schwarzschild solution. This was originally understood from the Hamiltonian point of view \cite{Regge:1974zd}, and more recently investigated for the Lagrangian description of the theory \cite{Mann:2005yr, Mann:2006bd, Mann:2008ay, Compere:2011db, Virmani:2011gh}.

In this paper we give a proper variational formulation for a theory of gravity coupled to a massive vector with ``Asymptotically Lifshitz'' solutions. We start by determining a set of boundary conditions for the fields that generalizes the Lifshitz asymptotics and includes spacetimes with curved spatial sections. We then determine the proper definition of the variational principle for the theory by supplementing the action \eqref{VectorFormulation} with an appropriate set of surface terms (boundary counterterms), a procedure that has come to be known as Holographic Renormalization \cite{Balasubramanian:1999re, Emparan:1999pm, de Boer:1999xf, de Haro:2000xn, Papadimitriou:2005ii, Mann:2005yr, Mann:2009id}. This extends the results of \cite{Ross:2009ar}, which investigated the variational formulation of theories with solutions that asymptote to \eqref{SimpleLifshitzMetric}-\eqref{SimpleLifshitzVector}.

There are two significant assumptions in our approach. First, a completely general analysis of the equations of motion is  quite difficult, so we make a few simplifying assumptions -- guided by the form of known solutions -- about the asymptotic behavior of the fields. While less general, this renders the analysis tractable, and highlights key features that must be present in a full treatment of the problem. Second, we focus on gravitational theories that provide a dual description of CM models with dynamical exponent $z=2$. Thus, we {\it do not} claim to have a completely general definition of an Asymptotically Lifshitz spacetime (even for the specific case $z=2$). While our results are quite broad, a more complete analysis of asymptotically Lifshitz boundary conditions (including time-dependence of the fields) is left for a future work.

The outline of our paper is as follows. In section \ref{sec:ALS} we formulate the equations of motion in a manner that lends itself to the analysis of spacetimes with anisotropic dependence on $r$, give boundary conditions that generalize the asymptotics of \eqref{SimpleLifshitzMetric}-\eqref{SimpleLifshitzVector} to solutions with curved spatial sections, and then construct solutions that satisfy these boundary conditions locally on a neighborhood of spatial infinity. We also describe the asymptotic symmetries of the theory. In section \ref{sec:VarPrin} we demonstrate that the action \eqref{VectorFormulation} is not suitable for the variational formulation of the theory, and then construct an action with the requisite properties. The construction is based on two requirements: that $\delta I=0$ on-shell for field variations having the same asymptotics as solutions to the equations of motion, and that $\delta I$ is finite on-shell for variations of boundary data that respect the kinematic constraints of the theory. Finiteness of the on-shell action follows from these more fundamental conditions. We then extend this construction and obtain an action such that $\delta I = 0$ on-shell for any field variations that preserve the boundary conditions of the theory. In section \ref{sec:BSTandCC} we use the procedure introduced by Hollands, Ishibashi, and Marolf (HIM) in \cite{Hollands:2005wt} to construct the ``improved'' boundary stress tensor and conserved charges for the theory. We also show how one of the actions introduced in section \ref{sec:VarPrin} allows the more familiar Brown-York (BY) stress tensor to be defined for the theory. The charges constructed from the BY stress tensor are also conserved, but only as a result of some fairly restrictive assumptions in our boundary conditions. The HIM charges are related in a suggestive way to the BY charges and a component of the boundary current dual to the massive vector field.   We explicitly demonstrate how to make use of our results in section \ref{sec:LTBH}, using Lifshitz topological black holes \cite{Mann:2009yx} as an example, and close our paper with a discussion in section \ref{sec:Discussion}.

Concerning notation, our calculations  involve two ADM-type decompositions: a $4 \to 3+1$ decomposition for studying the evolution of the fields along a spacelike coordinate $r$, followed by a $3 \to 2+1$ decomposition that splits the fields into parts with distinct asymptotic behavior. This requires notation for quantities on four-, three-, and two-dimensional spaces, which we summarize here. The spacetime is a four-dimensional manifold $\MM$ with metric $g_{\mu\nu}$ and coordinates $x^{\mu}$. A superscript `4' is prepended to spacetime curvatures and covariant derivatives, as in \eqref{Einstein}-\eqref{Proca}. Three-dimensional surfaces $M \subset \MM$ have coordinates $x^{a}$ and metric $h_{ab}$. The intrinsic curvatures on these surfaces are denoted $R_{ab}$ and $R$, the covariant derivative is $\nabla_{a}$, and the extrinsic curvature is $K_{ab}$. Finally, two-dimensional spatial surfaces $\Sigma \subset M$ have coordinates $x^{i}$ and metric $\sigma_{ij}$. The Ricci tensor on $\Sigma$ is $\RR_{ij}$, the covariant derivative is $D_{i}$, and the extrinsic curvature is $\theta_{ij}$.

\section{Asymptotically Lifshitz Spacetimes}
\label{sec:ALS}

In this section we generalize the asymptotics of the solution \eqref{SimpleLifshitzMetric}-\eqref{SimpleLifshitzVector} to spacetimes with curved spatial sections. We begin by expressing the equations of motion in a form suitable for studying these solutions. This is a two-step process, consisting of a $3+1$ split that emphasizes the evolution of the fields along the spacelike coordinate $r$, followed by a $2+1$ split that separates the fields into components with distinct asymptotic dependence on $r$. The resulting equations are used to identify a simple set of `asymptotically Lifshitz' boundary conditions, and then  solved in a neighborhood of spatial infinity to determine the asymptotic behavior of the fields.

\subsection{Decompositions of Fields and Equations of Motion}
\label{sec:Decompositions}

For the $3+1$ decomposition, we work with coordinates $x^{\mu} = (r,x^{a})$ adapted to a foliation $\{M_{r}\}$ of the spacetime by constant $r$ surfaces. It is convenient to partially fix the coordinate gauge so that the metric in the asymptotic region $r \gg \ell$ takes the form
\begin{gather}\label{4to3plus1}
  g_{\mu\nu}\,dx^{\mu} dx^{\nu} = \frac{\lsq}{r^2}\,dr^2 + h_{ab}(x^c,r)\,dx^a dx^b ~,
\end{gather}
with $h_{ab}(x^c,r)$ the metric on a surface $M_r$. Tensors are projected normal or parallel to $M_{r}$ by appropriate contractions with the unit vector $n^{\mu} = (r/\ell)\,\delta^{\mu}{}_{r}$ or the projector $P_{a}{}^{\mu} = \partial x^{\mu} / \partial x^{a}$. For the vector $A_{\mu}$ this gives a normal component $A_{n} = n^{\nu} A_{\nu}$ and a three-vector $A_{a} = P_{a}{}^{\mu} (A_{\mu})$, while the components of its field strength $F_{\mu\nu}$ are 
\begin{gather}\label{FieldStrengthPP}
	P_{a}{}^{\mu} P_{b}{}^{\nu} (F_{\mu\nu})  = \partial_{a} A_{b} - \partial_{b} A_{a} \defeq \FF_{ab} \\
	\label{Ba}
	P_{a}{}^{\nu}(n^{\mu} F_{\mu\nu}) = \Lie_{n} A_{a} - \partial_{a} A_{n} \defeq B_{a} ~,
\end{gather}
with $\Lie_{n}$ the Lie derivative along $n^{\mu}$. The projections of the equations of motion are carried out in the same manner, which gives five equations involving the fields $h_{ab}$, $A_{a}$, $A_{n}$, and their derivatives. The first three equations, obtained from the Einstein equations \eqref{Einstein}, are
\begin{align}
	\label{EinsteinNN}
	\frac{1}{2}\,\Big(\,K^2 - K^{ab}K_{ab}\,\Big) - \frac{1}{2}\,R =  & \,\, -\Lambda + \frac{1}{4}\,B^{a} B_{a} - \frac{1}{8}\,\FF^{ab} \FF_{ab} + \frac{m^2}{4}\,A_{n}{}^{2} - \frac{m^2}{4}\,A^{a} A_{a} \\ 
	\label{EinsteinNP}
	\nabla_{b} K^{ab} - \nabla^{a} K =  & \,\, \frac{m^2}{2}\,A_n \,A^{a} + \frac{1}{2}\,\FF^{ab} B_{b} \\	
 	\label{EinsteinPP}
		G_{ab} + h_{ab} \Lie_{n}K -  \Lie_{n}K_{ab} + \,&2\,K_{a}{}^{c} K_{bc} - K K_{ab} + \frac{1}{2}\,h_{ab}\,\Big(\,K^2 + K^{cd}K_{cd}\,\Big)   \\ \nonumber
		& = -h_{ab}\,\Lambda + \frac{1}{2}\,\FF_{a}{}^{c} \FF_{bc} - \frac{1}{8}\,h_{ab}\,\FF^{cd} \FF_{cd} + \frac{1}{2}\,B_{a} B_{b} - \frac{1}{4}\,h_{ab}\,B_{c} B^{c} \\ \nonumber 
		& \qquad + \frac{m^2}{4}\,A_a A_b - \frac{m^2}{4}\,h_{ab}\,A^{c} A_{c} - \frac{m^2}{4}\,h_{ab}\,A_{n}{}^{2} ~,
\end{align}
where 
\begin{equation}\label{Kdef}
K_{ab} = \frac{1}{2}\,\Lie_{n}h_{ab}
\end{equation}
is the extrinsic curvature of $M$, $K = h^{ab}\,K_{ab}$ is its trace, and $G_{ab}$ is the three-dimensional Einstein tensor. The two remaining equations are the projections of the Proca equations \eqref{Proca} 
\begin{gather} 
	\label{ProcaN}
  		- \nabla_{a} B^{a} = m^{2}\,A_{n} \\ 
	\label{ProcaP}
  		\nabla^{a} \FF_{ab} + \Lie_{n} B_{b} - 2\,B_{a} K^{a}{}_{b} + K B_{b} = m^{2} A_{b} ~. 
\end{gather}
In analogy with the usual ADM decomposition, \eqref{EinsteinNN}, \eqref{EinsteinNP}, and \eqref{ProcaN} are ``constraint equations'', while \eqref{EinsteinPP}, \eqref{ProcaP}, and the definitions (\ref{Ba}, \ref{Kdef}) of  $B_{a}$ and $K_{ab}$ are ``evolution equations''. These $3+1$ equations do not rely on any assumptions other than the choice of coordinate gauge.

The next step is a $2+1$ split of the three-dimensional coordinates into a time coordinate $t$ and spatial coordinates $x^{k}$, which gives a foliation $\{\Sigma_t\}$ of each $M_{r}$ by spatial surfaces of constant $t$. For a general choice of coordinates $x^{a}=(t,x^{k})$ the 3-metric takes the form
\begin{align}\label{General3Metric}
	h_{ab} dx^{a} dx^{b} = & \,\, -\alpha^{2}\,dt^2 + \sigma_{ij}\,(dx^i + \beta^{i}\,dt)
		(dx^j + \beta^{j}\, dt) ~,
\end{align}
where $\alpha$ is the lapse function, $\beta^{i}$ is the shift vector, and $\sigma_{ij}$ is the spatial 2-metric on a surface $\Sigma_{t}$. The 3-vector $A_{a}$ admits a similar split into a temporal component $\phi$ and a spatial 2-vector $\AA_i$
\begin{gather}\label{General3Vector}
  A_{a} dx^{a} = \phi\, dt + \AA_{i} 	\left(dx^i + \beta^{i} dt \right) ~.
\end{gather}
In principle, the $3+1$ equations \eqref{EinsteinNN} - \eqref{ProcaP} should now be projected normal and parallel to $\Sigma_t$ to yield a set of `$2+1+1$' equations of motion. However, implementing this for a completely general set of fields and solving the resulting equations is quite complicated, so we restrict our attention to fields whose asymptotic form is given by 
\begin{align}\label{MetricAssump}
	h_{ab} dx^{a} dx^{b} = & \,\, -\alpha^{2}(x^{k},r)\,dt^2 + \sigma_{ij}(x^{k},r)\,dx^{i} dx^{j} \\ \label{VectorAssump}
	A_{\mu} dx^{\mu} = & \,\, \phi(x^k, r)\,dt ~.
\end{align}
That is, for $r \gg \ell$ the fields are assumed to be independent of $t$, and certain components -- the shift vector $\beta^i$, the scalar $A_n$, and the 2-vector $\AA_i$ -- may be ignored. These restrictions let us avoid many of the complications of the general $2+1$ split, while still accommodating the original solution \eqref{SimpleLifshitzMetric}-\eqref{SimpleLifshitzVector}, the topological black hole spacetimes of \cite{Mann:2009yx}, and other solutions. They need not hold on the full spacetime $\MM$, and our results should apply equally well to more general field configurations when the deviations from \eqref{MetricAssump}-\eqref{VectorAssump} are suppressed in the asymptotic region.

The details of the $2+1$ decomposition of the equations of motion are discussed in appendix \ref{sec:3to2plus1}. The evolution equations associated with the definitions of $K_{ab}$ and $B_a$ split cleanly into temporal and spatial components
\begin{gather}
	K_{tt} = - \alpha \, \Lie_{n}\alpha \qquad K_{ti} = 0 \qquad K_{ij} = \frac{1}{2}\,\Lie_{n} \sigma_{ij} \\
	B_{t} = \Lie_{n} \phi \qquad B_{i} = 0 ~.
\end{gather}
The various projections of \eqref{EinsteinNN}-\eqref{ProcaP} yield a set of nine equations, four of which are automatically satisfied by fields of the form \eqref{MetricAssump}-\eqref{VectorAssump}. The remaining equations are
\begin{gather} \label{2plus1Eqn1}
	\frac{1}{2}\,\big(K^{i}{}_{i}\big)^2 - \frac{1}{2}\,K^{ij} K_{ij} + \frac{1}{\alpha}\,\Lie_{n}\alpha \, K^{i}{}_{i}- \frac{1}{2}\,\RR + \frac{1}{\alpha}\,D^{2} \alpha =  - \Lambda - \frac{1}{4\,\alpha^{2}}\,(\Lie_{n}\phi)^{2} + \frac{1}{4\,\alpha^{2}}\,D_i \phi D^i \phi + \frac{m^{2}}{4\,\alpha^{2}}\,\phi^{2}
\end{gather}
\begin{gather} \label{2plus1Eqn2}
	\frac{1}{2}\,\RR - \Lie_{n}K^{i}{}_{i} - \frac{1}{2}\,\big(K^{i}{}_{i}\big)^2 - \frac{1}{2}\,K^{ij} K_{ij} = \Lambda + \frac{1}{4\alpha^2}\,\big(\Lie_{n}\phi\big)^2 + \frac{1}{4\,\alpha^2}\,D_{i}\phi D^{i}\phi  + \frac{m^2}{4\alpha^2}\,\phi^{2} 
\end{gather}
\begin{gather} \label{2plus1Eqn3}
	D_i\Big(\,\frac{1}{\alpha}\,D^{i} \phi \,\Big) + \frac{1}{\alpha}\,\Lie_{n}{}^{2}\phi + \frac{1}{\alpha}\,\Lie_{n}\phi \, \Big( K^{i}{}_{i} - \frac{1}{\alpha}\,\Lie_{n}\alpha \Big) - \frac{m^2}{\alpha}\,\phi = 0 ~.
\end{gather}
\begin{gather} \label{2plus1Eqn4}
	D_j K^{ij} - D^i K^{j}{}_{j} = D^i\Big(\,\frac{1}{\alpha}\,\Lie_{n}\alpha\,\Big) - K^{ij}\,\frac{1}{\alpha}\,D_{j}\alpha + \frac{1}{2 \alpha^{2}}\,\Lie_{n}\alpha\,D^i\alpha - \frac{1}{2\alpha^{2}}\,\Lie_{n}\phi \,D^{i}\phi
\end{gather}
\begin{gather} \label{2plus1Eqn5}
\begin{split}
	\frac{1}{\alpha}\,\Big(\,\sigma_{ij} D^{2}\alpha & \, - D_i D_j \alpha \,\Big) + \sigma_{ij}\,\frac{1}{\alpha}\,\Lie_{n}{}^{2}\alpha + \sigma_{ij} \Lie_{n} K^{m}{}_{m} - \Lie_{n}K_{ij} + 2\,K_{i}{}^{m} K_{jm} - K^{m}{}_{m}\,K_{ij}  \\
	  + \,\, \frac{1}{\alpha}\,\Lie_{n}\alpha \,\big(& \,\sigma_{ij} K^{m}{}_{m} - K_{ij} \big) + \frac{1}{2}\,\sigma_{ij}\,\big( K^{m}{}_{m}\big)^2 + \frac{1}{2}\,\sigma_{ij}\,K^{mk} K_{mk}\\
	=& \,\, -\sigma_{ij}\,\Lambda - \frac{1}{2\,\alpha^{2}}\,D_i \phi D_j \phi + \frac{1}{4\,\alpha^{2}} \, \sigma_{ij} D_k \phi D^k \phi  + \frac{1}{4\,\alpha^{2}}\,\sigma_{ij} (\Lie_{n}\phi)^{2} + \frac{m^2}{4\,\alpha^{2}}\,\sigma_{ij}\,\phi^{2} ~,
		\end{split}
\end{gather}
where indices are lowered and raised using the 2-metric $\sigma_{ij}$ and its inverse. While the $3+1$ equations were completely general, these $2+1+1$ equations of motion are only valid when the fields have the restricted form \eqref{MetricAssump}-\eqref{VectorAssump}.

\subsection{Boundary Conditions and Asymptotic Behavior of Solutions}
\label{sec:BCandABS}

Boundary conditions that generalize the asymptotics of the solution \eqref{SimpleLifshitzMetric}-\eqref{SimpleLifshitzVector} can be determined from a straightforward analysis of the $2+1+1$ equations of motion. The leading behavior of the fields for $r \gg \ell$ is assumed to take the form
\begin{gather}\label{LeadingTerms}
	\sigma_{ij}(x^k,r) \sim \left(\frac{r}{\ell}\right)^{2} \sigma_{ij}^{\LO}(x^{k})  \qquad \alpha(x^k,r) \sim \left(\frac{r}{\ell}\right)^{z} \alpha^{\LO}(x^k) \qquad \phi(x^k,r) \sim \left(\frac{r}{\ell}\right)^{z} \phi^{\LO}(x^k) ~,
\end{gather}
where $\sigma_{ij}^{\LO}$, $\alpha^{\LO}$, and $\phi^{\LO}$ are smooth functions on $\Sigma_{t}$ that comprise the boundary data for the fields. Consistency with the equations of motion does not place any restrictions on $\sigma_{ij}^{\LO}$, but the functions $\alpha^{\LO}$ and $\phi^{\LO}$ must satisfy the constraint
\begin{gather}\label{BoundaryDataConstraint}
	\phi^{\LO} = \pm \,\alpha^{\LO}\,\sqrt{\frac{\,2\,(z-1)}{z}} ~.
\end{gather}
Invariance of the theory under $t \to -t$ means that the choice of sign in this equation is not important; we will always assume the positive sign. Thus, boundary data for the theory consists of a spatial 2-metric $\sigma_{ij}^{\LO}$ and a single scalar function that we define as $\alpha^{\LO} = e^{2\chi}$. The asymptotic behavior \eqref{LeadingTerms} with smooth boundary data that satisfies \eqref{BoundaryDataConstraint} will be referred to as asymptotically Lifshitz (AL) boundary conditions.

Solutions of the equations of motion with AL boundary conditions can be constructed locally on a neighborhood of spatial infinity for any choice of $\sigma_{ij}^{\LO}$ and $\chi$. There is no guarantee that these solutions also exist \textit{globally}, but we will only require properties of solutions that are present in the asymptotics.
First we perform a Taylor expansion of each field, starting from the leading behavior \eqref{LeadingTerms} 
\begin{gather}\label{TaylorExpansions}
	\sigma_{ij} = \sum_{n} \left(\frac{r}{\ell}\right)^{2-n} \sigma_{ij}^\ms{(n)}(x^k) \qquad\quad \alpha = \sum_{n}  \left(\frac{r}{\ell}\right)^{z-n} \alpha^\ms{(n)}(x^k) \qquad\quad \phi = \sum_{n}  \left(\frac{r}{\ell}\right)^{z-n} \phi^\ms{(n)}(x^k) ~.
\end{gather}
The equations of motion are then solved order-by-order in powers of $\ell/r$, which determines the coefficients in the expansions (as local functions of the boundary data) up to some finite order that depends on $z$. 
Beyond this point the coefficients also depend on dynamical aspects of the fields that are not fixed by the boundary conditions.

For the rest of this paper we will focus exclusively on theories with critical exponent $z=2$. In this case the expansions are qualitatively similar to an asymptotically AdS$_5$ spacetime \cite{de Haro:2000xn, Papadimitriou:2005ii}, with vanishing coefficients for the $n=1$ and $n=3$ terms in \eqref{TaylorExpansions}, and the coefficients of the $n=4$ terms only partially determined by the equations of motion. The asymptotic expansions of the fields out to this point are\,\footnote{For AL boundary conditions with $z=2$ the expansions do not contain $\log(r/\ell)$ terms. We have not ruled out such terms for different boundary conditions, or critical exponents $z > 2$.}
\begin{align} \label{SigmaExpansion}
	\sigma_{ij} = & \,\, \left(\frac{r}{\ell}\right)^{2} \, \left( \sigma_{ij}^{\LO} + \left(\frac{\ell}{r}\right)^{2} \sigma_{ij}^{\NLO} + \left(\frac{\ell}{r}\right)^{4} \sigma_{ij}^{\NNLO} + \ldots \right) \\ \label{LapseExpansion}
	\alpha = & \,\, \left(\frac{r}{\ell}\right)^{2} e^{2\,\chi} \left(1 + \left(\frac{\ell}{r}\right)^{2} \alpha^{\NLO} + \left(\frac{\ell}{r}\right)^{4} \alpha^{\NNLO} + \ldots \right) \\ \label{PhiExpansion}
	\phi = & \,\, \left(\frac{r}{\ell}\right)^{2} e^{2\,\chi} \left( 1 + \left(\frac{\ell}{r}\right)^{2} \phi^{\NLO} + \left(\frac{\ell}{r}\right)^{4} \phi^{\NNLO} + \ldots \right)~,
\end{align}
where we have pulled out a convenient overall factor of $e^{2\chi}$ in $\alpha$ and $\phi$, and use `$\ldots$' to represent higher-order terms ($n>4$) that will not be needed for our analysis. The $n=2$ and $n=4$ terms will be referred to as next-to-leading order (NLO) and next-to-next-to-leading order (NNLO), respectively. The equations of motion determine the NLO terms as two-derivative combinations of the boundary data 
\begin{gather} \label{sigmaNLO}
	\sigma_{ij}^{\NLO} = - \frac{\lsq}{8}\,\sigma_{ij}^{\LO}\,\RR^{\LO} + \lsq \, \DD_i \DD_j \chi + \lsq\,\DD_i \chi \DD_j \chi - \frac{\lsq}{4}\,\sigma_{ij}^{\LO}\,\DD^2 \chi - \frac{\lsq}{2}\,\sigma_{ij}^{\LO}\,\DD^k \chi \DD_k \chi \\ \label{LapseNLO}
	\alpha^{\NLO} = \frac{\lsq}{2} \, \DD_i \chi \DD^i \chi \\ \label{phiNLO}
	\phi^{\NLO} = \frac{\lsq}{8}\,\RR^{\LO} + \frac{\lsq}{4}\,\DD^{2} \chi + \frac{\lsq}{2}\,\DD_i \chi \DD^i \chi ~,
\end{gather}
where $\DD_{i}$ is the two-dimensional covariant derivative compatible with $\sigma_{ij}^{\LO}$ and $\RR^{\LO}$ is its scalar curvature. Thus, the NLO terms are fixed by the kinematicas of the theory. But at NNLO the fields are only partially determined by the boundary data. A single scalar degree of freedom appears, which we take to be the NNLO term $\phi^{\NNLO}$ in the expansion of the vector field. The NNLO term in the expansion of the lapse, $\alpha^{\NNLO}$, can be written in terms of $\phi^{\NNLO}$ and four-derivative combinations of the boundary data as
\begin{align} 	\label{LapseNNLO}
			\alpha^{\NNLO} = & \,\, - \frac{1}{5}\,\phi^{\NNLO} - \frac{\ell^{\,4}}{320}\,\big(\,\RR^{\LO}\,\big)^{2} - \frac{\ell^{\,4}}{80}\,\DD^{2}\RR^{\LO} - \frac{3\,\ell^{\,4}}{80}\,\DD_{i}\RR^{\LO}\,\DD^{i}\chi  + \frac{\ell^{\,4}}{80}\,\RR^{\LO}\,\DD^{i}\chi \DD_{i}\chi \\ \nonumber
			& \,\, + \frac{\ell^{\,4}}{16}\,\DD^{2}\chi \, \DD^{2}\chi + \frac{\ell^{\,4}}{40}\,\DD^{2}\chi\,\DD^{i}\chi \DD_{i}\chi
				- \frac{3\ell^{\,4}}{40}\,\DD^{i}\chi \,\DD^{2}\DD_{i}\chi - \frac{\ell^{\,4}}{40}\,\DD^{2}\DD^{2}\chi + \frac{3\,\ell^{\,4}}{40}\,\big(\,\DD^{i}\chi\DD_{i}\chi\,\big)^2 ~.
\end{align}
Both the trace and divergence of the NNLO term in the spatial metric are fixed by the equations of motion. Like $\alpha^{\NNLO}$, the trace $\sigma^{\NNLO} = \sigma^{ij}_{\LO}\,\sigma_{ij}^{\NNLO}$ depends on the scalar degree of freedom
\begin{align} \label{TrsigmaNNLO}
		\sigma^{\NNLO} = & \,\,	- \frac{2}{5}\,\phi^{\NNLO} + \frac{3 \ell^{\,4}}{320}\,\big(\,\RR^{\LO}\,\big)^{2} + \frac{\ell^{\,4}}{160}\,\DD^{2}\RR^{\LO} - \frac{\ell^{\,4}}{80}\,\DD_{i}\RR^{\LO}\,\DD^{i}\chi  
		- \frac{ \ell^{\,4}}{10}\,\RR^{\LO}\,\DD^{i}\chi \DD_{i}\chi \\ \nonumber
			& \,\, -\frac{\ell^{\,4}}{16}\,\DD^{2}\chi \, \DD^{2}\chi - \frac{9 \ell^{\,4}}{20}\,\DD^{2}\chi\,\DD^{i}\chi \DD_{i}\chi
				- \frac{\ell^{\,4}}{40}\,\DD^{i}\chi \,\DD^{2}\DD_{i}\chi + \frac{\ell^{\,4}}{80}\,\DD^{2}\DD^{2}\chi + \frac{3\,\ell^{\,4}}{20}\,\big(\,\DD^{i}\chi\DD_{i}\chi\,\big)^2 \\ \nonumber
			& \,\, + \frac{\ell^{\,4}}{2}\,\DD^i \chi \DD^j \chi \, \DD_i \DD_j \chi	+ \frac{\ell^{\,4}}{4}\,\DD^i \DD^j \chi \, \DD_i \DD_j \chi ~.
\end{align}
The divergence is determined by a complicated equation of the form
\begin{gather}\label{sigmaijNNLOEqn}
	\DD^{j}\sigma_{ij}^{\NNLO} + 2\,\sigma_{ij}^{\NNLO}\,\DD^{j}\chi + \mathscr{V}_i = 0 ~,
\end{gather}
where $\mathscr{V}_i$ is a combination of LO terms, NLO terms, and their derivatives. Solutions of this equation include a homogenous part annihilated by the first two terms, which may be written in terms of a transverse tensor $Y_{ij}$ as $\sigma_{ij}^{\NNLO \ts{hom}} = e^{-2\,\chi}\,Y_{ij}$. The full result for $\sigma_{ij}^{\NNLO}$ is
\begin{align} \label{sigmaijNNLO}
	\sigma_{ij}^{\NNLO} = & \,\, e^{-2\,\chi}\,Y_{ij} - \frac{1}{5}\,\sigma_{ij}^{\LO}\,\phi^{\NNLO} + \frac{3\,\ell^{\,4}}{640}\,\sigma_{ij}^{\LO}\,\big(\,\RR^{\LO}\,\big)^{2} - \frac{\,\ell^{\,4}}{32}\,\big(\, \DD_i \chi \DD_j \RR^{\LO} + \DD_i \RR^{\LO} \DD_j \chi\,\big) - \frac{\ell^{\,4}}{32}\,\DD_i \DD_j \RR^{\LO} \\ \nonumber
	& \,\, + \frac{3\,\ell^{\,4}}{160}\,\sigma_{ij}^{\LO}\,\DD^{2} \RR^{\LO} - \frac{\ell^{\,4}}{160}\,\sigma_{ij}^{\LO}\,\RR^{\LO}\,\DD^{2}\chi - \frac{\ell^{\,4}}{4}\,\DD_i \chi \DD_j \chi \, \DD^{2}\chi
	- \frac{\ell^{\,4}}{8}\,\big(\,\DD_i \chi \,\DD^2 \DD_j \chi + \DD_j \chi \,\DD^2 \DD_i \chi\,\big) \\ \nonumber
	& \,\, - \frac{\ell^{\,4}}{16}\,\DD^{2} \DD_i \DD_j \chi 
	  + \frac{\ell^{\,4}}{40}\,\sigma_{ij}^{\LO}\,\DD^{k}\chi\,\DD_{k}\RR^{\LO} 
	  - \frac{7\ell^{\,4}}{160}\,\sigma_{ij}^{\LO}\,\RR^{\LO}\,\DD^{k}\chi \, \DD_{k}\chi
	  - \frac{\ell^{\,4}}{4}\,\DD_i \DD_j \chi \, \DD^{k}\chi \DD_{k}\chi \\ \nonumber
	& \,\, + \frac{\ell^{\,4}}{4}\,\big(\,\DD_i \DD_k \chi \, \DD_j \chi + \DD_j \DD_k \chi \, \DD_i \chi \,\big)\,\DD^k \chi 
	  - \frac{\ell^{\,4}}{32}\,\sigma_{ij}^{\LO}\,\big(\,\DD^{2}\chi\,\big)^{2}
	  + \frac{\ell^{\,4}}{40}\,\sigma_{ij}^{\LO}\,\DD^{2}\chi \, \DD^k \chi \DD_k \chi \\ \nonumber
	& \,\, + \frac{9\,\ell^{\,4}}{80}\,\sigma_{ij}^{\LO}\,\DD^{k}\chi \, \DD^{2} \DD_{k} \chi
	  + \frac{3\,\ell^{\,4}}{80}\,\sigma_{ij}^{\LO}\,\DD^{2} \DD^{2} \chi 
	  + \frac{3\,\ell^{\,4}}{40}\,\sigma_{ij}^{\LO}\,\big(\,\DD^k \chi \, \DD_k \chi \,\big)^2
	  + \frac{\ell^{\,4}}{8}\,\sigma_{ij}^{\LO}\,\DD^k \DD^l \chi \, \DD_k \DD_l \chi ~.
\end{align}
The trace of the inhomogenous part of this solution agrees with \eqref{TrsigmaNNLO}, so in addition to being transverse the tensor $Y_{ij}$ must also be traceless. We show in appendix \ref{sec:TTBD} that symmetric, transverse-traceless tensors on $\Sigma_t$ built from four-derivative combinations of the boundary data vanish identically, which confirms that $Y_{ij}$ is independent of the boundary data.

The expressions for the fields out to NNLO are quite complicated, so it is useful to look at a special case where the results simplify. In section 5 we consider the topological black hole solutions  \cite{Mann:2009yx}, which  are examples of AL fields where the function $\chi$ -- the boundary data for $\alpha$ and $\phi$ -- is constant. 
For simplicity we will take $\chi = 0$, in which case the coefficients in the asymptotic expansions \eqref{SigmaExpansion}-\eqref{PhiExpansion} are
\begin{gather} \label{SimpleSigma}
	\sigma_{ij}^{\NLO} = - \frac{\lsq}{8}\,\sigma_{ij}^{\LO}\,\RR^{\LO} \qquad \sigma_{ij}^{\NNLO} =  \,Y_{ij} - \frac{1}{5}\,\sigma_{ij}^{\LO}\,\phi^{\NNLO} + \frac{3\,\ell^{\,4}}{640}\,\sigma_{ij}^{\LO}\,\big(\,\RR^{\LO}\,\big)^{2}  - \frac{\ell^{\,4}}{32}\,\DD_i \DD_j \RR^{\LO} + \frac{3\,\ell^{\,4}}{160}\,\sigma_{ij}^{\LO}\,\DD^{2} \RR^{\LO}\\ \label{SimpleLapse}
	\alpha^{\NLO} = 0 \qquad \alpha^{\NNLO} = - \frac{1}{5}\,\phi^{\NNLO} - \frac{\ell^{\,4}}{320}\,\big(\,\RR^{\LO}\,\big)^{2} - \frac{\ell^{\,4}}{80}\,\DD^{2}\RR^{\LO} \\ \label{SimplePhi}
	\phi^{\NLO} = \frac{\lsq}{8}\,\RR^{\LO}~.
\end{gather}
As with the more general solution, the degrees of freedom appearing at NNLO are a scalar $\phi^{\NNLO}$ and a transverse-traceless tensor $Y_{ij}$.

To summarize, AL boundary conditions for the restricted set of fields \eqref{MetricAssump}-\eqref{VectorAssump} are given by the leading asymptotic behavior \eqref{LeadingTerms}, with boundary data consisting of a scalar function and a metric on some two-dimensional spatial surface $\Sigma_t$. Solutions of the equations of motion can be constructed locally on a neighborhood of spatial infinity, and take the form \eqref{SigmaExpansion} - \eqref{PhiExpansion} for theories with critical exponent $z=2$. The boundary data at LO fixes the NLO terms according to \eqref{sigmaNLO} - \eqref{phiNLO}, but the NNLO terms are only determined up to a scalar degree of freedom and the transverse-traceless part of the spatial metric. At higher orders in the expansion, the coefficients depend on this dynamical information in addition to the boundary data.

\subsection{Asymptotic Symmetries}
\label{sec:AS}

The asymptotic symmetries of a theory are the diffeomorphisms that act on the fields at spatial infinity, while preserving the boundary conditions. For a theory with boundary conditions that admit the Lifshitz solution \eqref{SimpleLifshitzMetric}-\eqref{SimpleLifshitzVector}, these include the non-relativistic scaling transformations \eqref{LifshitzScaling}. For other choices of AL boundary conditions the Lifshitz scaling transformations are typically broken, though other symmetries may exist.

Under an infinitesimal coordinate transformation $x^{\mu} \to x^{\mu} - \xi^{\mu}(x)$ the metric and vector transform as
\begin{gather}\label{Diffeos}
	\delta_{\xi} g_{\mu\nu} = \Lie_{\xi} g_{\mu\nu} \qquad \delta_{\xi} A_{\mu} = \Lie_{\xi} A_{\mu} ~.
\end{gather}
In order to preserve the coordinate gauge \eqref{4to3plus1}, these diffeomorphisms must take the general form
\begin{gather}\label{PreserveCoordGauge}
	\xi^{r} = r\,\gamma(x^{c}) \qquad \xi^{a} = \xi^{a}_{\LO}(x^c) - \int dr \frac{\,\ell^{2}}{r}\,h^{ab}\,\partial_{b} \gamma ~,
\end{gather}
where $\gamma$ and $\xi^{a}_{\LO}$ may depend on the three-dimensional coordinates $x^c$, but not on $r$. Requiring that the diffeomorphisms also respect the asymptotic form of the fields \eqref{MetricAssump}-\eqref{VectorAssump} further restricts the coordinate-dependence of the components of $\xi^{\mu}$, and we find
\begin{gather}\label{Diffeo2}
	\xi^{r} = r\,\gamma(x^{k}) \qquad \xi^{t} = \xi^{t}_{\LO}(t) \qquad \xi^{i} = \eps^{i}_{\LO}(x^{k}) + \left(\frac{\ell}{r}\right)^{2}\,\eps^{i}_{\NLO}(x^{k}) + \left(\frac{\ell}{r}\right)^{4}\,\eps^{i}_{\NNLO}(x^{k}) + \ldots ~.
\end{gather}
The NLO and NNLO terms in $\xi^i$ are obtained using the asymptotic expansion of $\sigma^{ij}$ in the integral in \eqref{PreserveCoordGauge}, which gives
\begin{gather}
	\eps^{i}_{\NLO} = \frac{\ell^{2}}{2}\,\DD^{i} \gamma \qquad 	\eps^{i}_{\NNLO} = - \frac{\ell^{2}}{4}\,\sigma^{ij}_{\NLO}\DD_{j} \gamma ~.
\end{gather}
Thus, diffeomorphisms that preserve both the coordinate gauge and the asymptotic form of the fields consist of a rescaling of $r$ that depends on the spatial coordinates $x^k$, a $t$-dependent reparameterization of time, and a diffeomorphism on $\Sigma_t$ that includes contributions that are sub-leading in $r$. 

The action of the diffeomorphisms on the fields can be worked out from the transformations \eqref{Diffeos}. We are primarily concerned with the response of the fields at LO and NLO in the asymptotic expansion, since these terms are completely fixed by boundary conditions and kinematics. Under the diffeomorphism \eqref{Diffeo2}, the LO and NLO terms in the expansion \eqref{SigmaExpansion} for the spatial metric transform as
\begin{align} 
	\label{SigmaTransformationLO}
	\delta_{\xi} \sigma_{ij}^{\LO} = & \,\, 2\,\gamma\,\sigma_{ij}^{\LO} + \Lie_{\eps_{\LO}} \sigma_{ij}^{\LO} \\
	\label{SigmaTransformationNLO}
	\delta_{\xi} \sigma_{ij}^{\NLO} = & \,\, \Lie_{\eps_{\LO}} \sigma_{ij}^{\NLO} + \Lie_{\eps_{\NLO}} \sigma_{ij}^{\LO} ~.
\end{align}
Now consider the fields $\alpha$ and $\phi$. Before the constraint \eqref{BoundaryDataConstraint} is applied, the asymptotic expansions take the form
\begin{align}\label{Expansion2}
	\alpha = \left(\frac{r}{\ell}\right)^{2} \alpha^{\LO}\, \left( 1 + \left(\frac{\ell}{r}\right)^{2} \alpha^{\NLO} + \ldots \right) \qquad \qquad
	\phi = \left(\frac{r}{\ell}\right)^{2} \phi^{\LO}\, \left( 1 + \left(\frac{\ell}{r}\right)^{2} \phi^{\NLO} + \ldots \right) ~.
\end{align}
The terms in the expansion of the lapse transform according to
\begin{align}\label{LapseTransformationLO}
	\delta_{\xi}\alpha^{\LO} = & \,\, \left( 2\,\gamma + \partial_{t}\,\xi^{t}_{\LO}\right)\alpha^{\LO} + \Lie_{\eps_{\LO}} \alpha^{\LO} \\ \label{LapseTransformationNLO}
	\delta_{\xi}\alpha^{\NLO} = & \,\, -2 \,\gamma\, \alpha^{\NLO} + \Lie_{\eps_{\LO}} \alpha^{\NLO} + \Lie_{\eps_{\NLO}} \log \alpha^{\LO} ~,
\end{align}
and likewise for the terms in the expansion of $\phi$
\begin{align}\label{PhiTransformationLO} 
	\delta_{\xi}\phi^{\LO} = & \,\, \left( 2\,\gamma + \partial_{t}\,\xi^{t}_{\LO}\right)\phi^{\LO} + \Lie_{\eps_{\LO}} \phi^{\LO} \\ \label{PhiTransformationNLO}
	\delta_{\xi}\phi^{\NLO} = & \,\, -2 \,\gamma\, \alpha^{\NLO} + \Lie_{\eps_{\LO}} \phi^{\NLO} + \Lie_{\eps_{\NLO}} \log \phi^{\LO} ~.
\end{align}
The overall factors of $\alpha^{\LO}$ and $\phi^{\LO}$ in \eqref{Expansion2} result in slightly non-standard transformations for the NLO terms. Notice that the transformations of the LO terms preserve the constraint on the boundary data \eqref{BoundaryDataConstraint}, so we can restrict our attention to $\alpha^{\LO}$ and $\sigma_{ij}^{\LO}$ when determining the diffeomorphisms that preserve the boundary conditions.

Asymptotic symmetries act at spatial infinity but preserve the boundary conditions, which means that the transformations of the LO terms in the fields must vanish. In the case of the lapse, requiring $\delta_{\xi} \alpha^{\LO} = 0$ immediately restricts the possible form of the time reparameterization. Since $\gamma$ and $\alpha^{\LO}$ are functions of $x^{k}$, it follows that $\partial_t \xi^{t}_{\LO}$ cannot depend on $t$ if \eqref{LapseTransformationLO} is to vanish, and the only possibilities are constant rescalings or constant translations
\begin{gather}
	\xi^{t}_{\LO} = -2\,\lambda\,t - \delta t ~.
\end{gather}
Taking this into account, the conditions for the diffeomorphism to preserve the boundary data can be written as
\begin{gather}
	\label{alphaAsymptoticSymmetries}
	\Lie_{\eps_{\LO}} \alpha^{\LO} = \left( 2\,\lambda + \DD_{k} \eps^{k}_{\LO} \right) \alpha^{\LO} \\
	\label{sigmaAsymptoticSymmetries}
	\Lie_{\eps_{\LO}} \sigma_{ij}^{\LO} = \sigma_{ij}^{\LO} \,\DD_k \eps^{k}_{\LO} ~,
\end{gather}
where we have used the trace of \eqref{SigmaTransformationLO} to express $\gamma$ in terms of the divergence $\DD_k \xi^{k}_{\LO}$. For given boundary data $\alpha^{\LO}$ and $\sigma_{ij}^{\LO}$, asymptotic symmetries are associated with $\lambda$ and $\xi^{i}_{\LO}$ that satisfy these equations. For the original Lifshitz solution \eqref{SimpleLifshitzMetric}-\eqref{SimpleLifshitzVector} we have $\chi = 0$ and $\sigma_{ij}^{\LO} = \delta_{ij}$, and in that case the asymptotic symmetries include the Lifshitz scaling generated by
\begin{gather}
	\xi^{r} = \lambda\,r \qquad \xi^{t} = -2\,\lambda\,t \qquad \xi^{i} = -\lambda\,x^{i} ~.
\end{gather}
Other choices of boundary data may or may not allow non-trivial solutions of \eqref{alphaAsymptoticSymmetries}-\eqref{sigmaAsymptoticSymmetries}. Notice, however, that the asymptotic symmetries always include the constant time translation $t \to t + \delta t$.

One application of our results is to AL solutions with $\alpha^{\LO}=\phi^{\LO}=1$ and spatial sections of constant (non-zero) scalar curvature. In these cases, the asymptotic symmetries do not include Lifshitz scaling transformations. The easiest way to demonstrate this is to consider the transformation of the NLO terms in the fields. Since the NLO terms are completely fixed by the kinematics, they should not change if the diffeomorphism preserves the boundary conditions. Using the $\chi =0$ solutions  \eqref{SimpleSigma} - \eqref{SimplePhi}, the transformations of the fields at NLO become
\begin{align}
	\delta_{\xi} \alpha^{\NLO} = & \,\, 0 \\
	\delta_{\xi} \phi^{\NLO} = & \,\, \frac{\ell^{2}}{8}\,\left(\eps^{k}_{\LO} \partial_k - 2\,\lambda \right) \RR^{\LO} \\
	\delta_{\xi} \sigma_{ij}^{\NLO} = & \,\, - \frac{\ell^{2}}{8}\,\sigma_{ij}^{\LO}\,\left(\eps^{k}_{\LO} \partial_k - 2\,\lambda \right) \RR^{\LO} ~,
\end{align}
where we have simplified some terms and canceled others using $\delta_{\xi}\alpha^{\LO} = 0$ and $\delta_{\xi} \sigma_{ij}^{\LO} = 0$. If the scalar curvature $\RR^{\LO}$ is a non-zero constant, then the NLO terms in the vector and spatial metric only vanish if $\lambda = 0$. The equations \eqref{alphaAsymptoticSymmetries}-\eqref{sigmaAsymptoticSymmetries} might have non-trivial solutions in these cases, but they do not include the Lifshitz scaling transformations \eqref{LifshitzScaling}.

\section{A Variational Principle}
\label{sec:VarPrin}

A variational principle identifies solutions of a theory as stationary points of an action. Generally speaking, there is some space of allowed field configurations, and the action must be stationary for any variation of the fields within the space. A basic requirement for a ``well-defined'' variational principle is that this space should include generic field configurations with the same asymptotic behavior as any physically reasonable solution of the theory \cite{Regge:1974zd}. If a proposed action is not stationary for arbitrary variations within this space, then it is not suitable for the variational formulation of the theory.  Of course, one may have additional applications of the action in mind, which require that it have certain properties on an even larger space of field configurations.
 

The results of the last section can be used to make the basic requirement on the space of field configurations more precise for theories with AL boundary conditions and critical exponent $z=2$.
In that case, all solutions are identical at LO and NLO in the asymptotic expansion, but they may differ at NNLO. This means that an action -- when evaluated on a solution of the equations of motion -- must be stationary for independent variations of the fields at NNLO. As we pointed out in the introduction, the action \eqref{VectorFormulation} does not have this property. The response of that action to a small change in the fields is 
\begin{gather}\label{OldActionVariation}
	\delta I = \frac{1}{2 \kappa^2} \int_{\MM} \nts \dd^{4}x \,\sqrt{-g}\,\left(\vphantom{\pi^{ab}}\,\EE^{\mu\nu}\,\delta g_{\mu\nu} + \EE^{\mu}\,\delta A_{\mu} \right) + \frac{1}{\kappa^2}\int_{\dM} \bns \dd^{3}x\,\sqrt{-h}\,\left(\,\pi^{ab}\,\delta h_{ab} + \pi^{a}\,\delta A_{a} \right) ~,
\end{gather}
where $\EE^{\mu\nu}=0$ and $\EE^{\mu}=0$ give the equations of motion \eqref{Einstein}-\eqref{Proca}, and the coefficients of the field variations in the surface integral are
\begin{gather}
	\pi^{ab} = \frac{1}{2}\,\left(h^{ab}\,K - K^{ab} \right) \qquad \pi^{a} = - \frac{1}{2} \, n_{\mu}\,F^{\mu a} ~.
\end{gather}
The bulk integral in \eqref{OldActionVariation} vanishes for solutions of the equations of motion, but the surface integral does not. To see why this is the case, it is convenient to work with the fields $\alpha$, $\sigma_{ij}$, and $\phi$. The surface integral is now given by
\begin{gather}\label{OldActionVariation2}
	\delta I\,\big|_\ms{\EE=0} = \frac{1}{\kappa^2} \int_{\dM} \bns \dd^{3}x\,\sqrt{-h}\,\left(-2\,\alpha\,\pi^{tt}\,\delta \alpha + \pi^{ij}\,\delta \sigma_{ij} + \pi^{t}\,\delta \phi \right) ~.
\end{gather}
To make sense of this integral, which contains factors that either vanish or diverge at spatial infinity, the integrand should be evaluated on a regulating surface $M_r$, with $r \gg \ell$. The asymptotic expansions for the fields can then be used to determine the $r \to \infty$ limit. The field variations fall off as $r^{-2}$ in this limit\,\footnote{This is not the case if we work with $h_{tt}$ instead of $\alpha$. The variation $\delta h_{tt} = -2 \,\alpha \,\delta \alpha$ approaches a constant at spatial infinity.}, since they behave like NNLO terms in the asymptotic expansions \eqref{SigmaExpansion}-\eqref{PhiExpansion}. But their coefficients -- including the contribution from the volume factor -- grow as $r^2$. As a result, the surface integral makes a finite but non-zero contribution to $\delta I$ as $M_r$ is taken to spatial infinity\,\footnote{Notice that the field variations themselves go to zero in this limit. A Dirichlet boundary value problem is not equivalent, in this case, to a well-defined variational principle.}. We conclude that the action \eqref{VectorFormulation} is not stationary under the full class of variations required for a well-defined variational principle. 

\subsection{A Minimal Action}
\label{sec:MinimalAction}

An action with the appropriate variational properties is obtained by adding new surface terms to \eqref{VectorFormulation}. Since the action is a functional of fields that diverge at spatial infinity, we work on a compact region $\MM_r \subset \MM$ bounded by a surface $M_r$ of constant $r \gg \ell$. The action is then defined as the $r \to \infty$ limit of a functional $I_r$ on this cut-off spacetime. We assume that the new surface terms are at most quadratic in the fields and their derivatives, so $I_r$ is given by
\begin{align}\label{ActionWithCT}
		I_r = & \,\, \frac{1}{2\kappa^{2}} \int_{\MM_r} \nts \textrm{d}^{4}x \, \sqrt{-g} \left( \RS - 2\,\Lambda - \frac{1}{4}\,F^{\mu\nu}\,F_{\mu\nu} - \frac{m^2}{2}\,A^{\mu} A_{\mu}\right) 
		+ \frac{1}{\kappa^2} \int_{M_r} \nts \textrm{d}^{3}x\,\sqrt{-h}\,K \\ \nonumber
		& \,\, + \frac{1}{\kappa^{2}} \int_{M_r} \nts \textrm{d}^{3}x\,\sqrt{-h}\, \left( c_{0} + c_{1}\,A^{a} A_{a} + c_{2}\,R + c_{3}\,\FF^{ab} \FF_{ab} \right) ~.
\end{align}
Any calculation involving the action or its variation is performed using this functional, and completed by the $r \to \infty$ limit that takes $M_r$ to spatial infinity. Other procedures for cutting off the spacetime may give different results, so our choice of limiting procedure should be thought of as part of the definition of the theory.

The coefficients of the surface terms in \eqref{ActionWithCT} are fixed by demanding a specific response from the action for two different types of field variations. First, $\delta I$ must vanish on-shell for small, independent variations of the fields at NNLO. This guarantees that solutions of the equations of motion are stationary points of the action within the basic space of field configurations needed for a well-defined variational principle. Second, the action should have a finite response to field variations caused by small changes in the boundary data that respect the constraint \eqref{BoundaryDataConstraint}. This insures that both the on-shell action and the conserved charges (obtained from a suitably defined boundary stress tensor) are finite. It also has the effect of enlarging the space of field configurations allowed by the variational principle. An action that meets these two requirements will be referred to as `minimal'. It is worth pointing out that the choice of surface terms in \eqref{ActionWithCT} is not unique, and other combinations of surface terms can be used to obtain a minimal action. This issue will be discussed in more detail at the end of this section.

We will now use the conditions on the variation of the action to determine the coefficients $\{c_i\}$ in \eqref{ActionWithCT}. Since the new surface terms are intrinsic to the regulating surface $M_r$, they do not introduce additional bulk terms in $\delta I$. The on-shell variation of the action has the same basic form as before
\begin{gather}\label{NewActionVariation}
	\delta I\,\big|_\ms{\EE=0} = \frac{1}{\kappa^{2}} \int_{M_r} \nts \dd^{3}x \sqrt{-h}\,\left[ \raisebox{13pt}{} \left(\pi^{ab} + p^{ab}\right)\,\delta h_{ab} + \left( \pi^{a} + p^{a} \right)\,\delta A_{a} \right] ~,
\end{gather}
with the contributions from the new surface terms given by
\begin{align}
\label{CTmetvar}
p^{ab} = & \,\, \frac{1}{2}\,c_0 \,h^{ab} + c_1 \left(\frac{1}{2}\,h^{ab}\,A^c A_c - A^a A^b \right) + c_2 \left( \frac{1}{2}\,h^{ab}\,R - R^{ab}\right) + c_3 \left(\frac{1}{2}\,h^{ab}\,\FF^{cd}\FF_{cd} -2\,\FF^{a}{}_{c}\, \FF^{bc}\right) \\ \label{CTvectvar}
p^{a} = & \,\, 2\,c_1\,A^a - 4\,c_3\,\nabla_{b} F^{ba} ~.
\end{align}
For the variational principle to be well-defined, \eqref{NewActionVariation} should vanish for independent variations of the fields at NNLO. These can be written as
\begin{gather}\label{NNLOvariations}
	\delta \sigma_{ij} = \left(\frac{\ell}{r}\right)^{2}\,\delta \sigma_{ij}^{\NNLO} \qquad 
	\delta \,\alpha = \left(\frac{\ell}{r}\right)^{2}\,e^{2\,\chi}\,\delta \alpha^{\NNLO} \qquad
	\delta \, \phi =  \left( \frac{\ell}{r} \right)^{2} e^{2\,\chi} \,\delta  \phi^{\NNLO} ~,
\end{gather}
with the factors of $e^{2\chi}$ in $\delta \alpha$ and $\delta \phi$ included for convenience. Using these expressions in $\delta I$, we have
\begin{align}\label{DynamicalActionVariation}
	\delta I\,\big|_\ms{\EE=0} = \frac{1}{\kappa^{2}} \int_{M_r} \nts \dd^{3}x \,\sqrt{-h}\, \bigg[ \,& \big(\pi^{ij} + p^{ij}\big) \left(\frac{\ell}{r}\right)^{2} \delta \sigma_{ij}^{\NNLO} -2\, \alpha\, \big( \pi^{tt} + p^{tt}\big)\left(\frac{\ell}{r}\right)^{2} e^{2\chi}\,\delta \alpha^{\NNLO} \\ \nonumber
	& + \big( \pi^{t} + p^{t} \big)\left(\frac{\ell}{r}\right)^{2} e^{2\,\chi} \,\delta  \phi^{\NNLO} \,\bigg] ~.
\end{align}
For $r \gg \ell$, the $r^{-2}$ behavior of each field variation is canceled by the $r^2$ growth of its coefficient, leaving three independent terms in the surface integral that are finite and non-zero as $r \to \infty$. For the action to be stationary, the coefficients $\{c_i\}$ must be tuned to cancel these finite contributions. Working out the asymptotic expansions for each term in the integrand, the leading behavior is
\begin{align}
	\sqrt{-h}\,\big( \pi^{ij} + p^{ij} \big)\,\left(\frac{\ell}{r}\right)^{2}\,\delta \sigma_{ij}^{\NNLO} = & \,\, \sqrt{\sigma^{\LO}}\,e^{2 \chi}\,\left( \frac{3}{2\,\ell} + \frac{c_0}{2} - \frac{c_1}{2} \right)\,\sigma^{ij}_{\LO} \, \delta \sigma_{ij}^{\NNLO} + \ldots \\
	-2\,\sqrt{-h}\,\alpha\,\big( \pi^{tt} + p^{tt} \big)\left(\frac{\ell}{r}\right)^{2} e^{2 \chi}\,\delta \alpha^{\NNLO} = & \,\, - \sqrt{\sigma^{\LO}}\,e^{2 \chi}\bigg( \frac{1}{\ell} +\frac{c_0}{2} + \frac{c_{1}}{2} \bigg) \,\delta \alpha^{\NNLO} + \ldots ~ \\
	\sqrt{-h}\,\big(\pi^{t} + p^{t} \big) \left(\frac{\ell}{r}\right)^{2} e^{2\,\chi} \,\delta  \phi^{\NNLO} = & \,\, \sqrt{\sigma^{\LO}}\,e^{2 \chi}\,\bigg( \frac{1}{\ell} - 2\,c_1 \bigg) \delta \phi^{\NNLO} + \ldots ~,
\end{align}
with `$\ldots$' indicating terms that vanish as $r \to \infty$. Only the first two surface terms that were added to \eqref{ActionWithCT} make finite contributions to $\delta I$ for the field variations \eqref{NNLOvariations}; the other terms fall off too rapidly at large $r$. This poses a potential problem, since there are two coefficients and three independent terms that must be addressed. However, all three terms are canceled by setting 
\begin{gather}
	c_0 = - \frac{5}{2\,\ell} \quad \quad c_1 = \frac{1}{2\,\ell} ~.
\end{gather}
Thus, requiring the on-shell action to be stationary for independent variations of the fields at NNLO determines the coefficients of two of the surface terms in the action.

The coefficients of the remaining surface terms are fixed by demanding a finite response of the action to Hamilton-Jacobi variations of the fields; i.e., field variations associated with small changes of the boundary data that satisfy the constraints of the theory. These variations take the form
\begin{gather}\label{HJVariations}
	\delta \sigma_{ij} =  \left( \frac{r}{\ell} \right)^2 \, \delta \sigma_{ij}^{\LO} + \ldots \qquad 
	\delta \alpha =  2\,  \left( \frac{r}{\ell} \right)^2 \, e^{2\,\chi}\,\delta \chi + \ldots \qquad
	\delta \phi =  2\,  \left( \frac{r}{\ell} \right)^2 \, e^{2\,\chi}\,\delta \chi + \ldots ~.
\end{gather}
Notice that the variations of $\alpha$ and $\phi$ have the same leading behavior, since the kinematic constraint \eqref{BoundaryDataConstraint} requires $\alpha^{\LO} = \phi^{\LO}$ when $z=2$. The `$\ldots$' in each expression is a reminder that the NLO terms \eqref{LapseNLO} - \eqref{sigmaNLO} will also change when the boundary data is varied. However, the contributions to $\delta I$ from the NLO terms in the field variations will vanish in the $r \to \infty$ limit, so we may ignore them in this calculation. The change in the action due to a Hamilton-Jacobi variation of the fields is
\begin{gather}\label{HJActionVariation}
	\delta I\,\big|_\ms{\EE=0} = \frac{1}{\kappa^{2}} \int_{M_r} \nts \dd^{3}x \,\sqrt{-h}\, \left[ \big(\pi^{ij} + p^{ij}\big) \left(\frac{r}{\ell}\right)^{2} \delta \sigma_{ij}^{\LO} + \bigg( - 4 \,\alpha \,\big( \pi^{tt} + p^{tt}\big) + 2\,\big( \pi^{t} + p^{t} \big)\bigg) \,\left(\frac{r}{\ell}\right)^{2} e^{2\chi}\,\delta \chi \, \right] ~.
\end{gather}
Working out the asymptotic expansion of the terms in the integrand reveals contributions proportional to $r^{2}$, which diverge as $r \to \infty$, and contributions independent of $r$, which are finite in that limit. The terms proportional to $r^{2}$ can be removed by tuning the coefficients of the remaining surface terms. First we consider the term in \eqref{HJActionVariation} proportional to $\delta \sigma_{ij}^{\LO}$:
\begin{align}
	\sqrt{-h}\,\big(\pi^{ij} + p^{ij}\big) \left(\frac{r}{\ell}\right)^{2} \delta \sigma_{ij}^{\LO} =  - \frac{1}{4}\,\left(\frac{r}{\ell}\right)^2\,e^{2\chi}\,\sqrt{\sigma^{\LO}}\,& \bigg[ \big( - 16\,c_2 -32\,c_3 -2\,\ell \big) \, \big( \DD^i \chi \DD^j \chi - \sigma^{ij}_{\LO}\,\DD^k \chi \DD_k \chi \big) \\ \nonumber
	& \quad + \big( - 8\,c_2 - 2\,\ell \big) \, \big( \DD^i \DD^j \chi - \sigma^{ij}_{\LO}\,\DD^k \DD_k \chi\big) \\ \nonumber
	& \quad + \big( \ell - 16\,c_3\big) \, \sigma^{ij}_{\LO}\,\DD^k \chi \DD_{k} \chi \, \bigg] \, \delta \sigma_{ij}^{\LO} + \ldots
\end{align}
with `$\ldots$' indicating the part that is finite as $r \to \infty$. The $r^2$ terms are canceled by setting 
\begin{gather}
	c_2 = - \frac{\ell}{4} \qquad c_3 = \frac{\ell}{16} ~.
\end{gather}
These values of $c_2$ and $c_3$ also cancel divergences in the $r^2$ terms proportional to $\delta \chi$, which are given by
\begin{align}\label{LOdeltaChiTerms}
\sqrt{-h} \,\bigg( - 4 \,\alpha \,& \big( \pi^{tt} + p^{tt}\big) + 2\,\big( \pi^{t} + p^{t} \big)\,\bigg) \, \left(\frac{r}{\ell}\right)^{2} e^{2\chi}\,\delta \chi =  \\ \nonumber
&\,\,  = \left(\frac{r}{\ell}\right)^{2} e^{2\chi} \,\sqrt{\sigma^{\LO}} \left[ \, \frac{1}{2}\,\big(4\,c_2 + \ell \, \big) \, \RR^{\LO} + \big( 16\,c_3 - \ell \, \big)\,\big(\DD^{k}\DD_{k}\chi + \DD^{k}\chi \DD_{k}\chi \big)\right]\,\delta\chi	+ \ldots ~.
\end{align}
Therfore, the on-shell response of the action to Hamilton-Jacobi variations of the fields is finite as $ r\to \infty$ if the coefficients of the surface terms are given by
\begin{gather}
	c_0 = - \frac{5}{2\,\ell} \qquad c_1 = \frac{1}{2\,\ell} \qquad c_2 = - \frac{\ell}{4} \qquad c_3 = \frac{\ell}{16} ~.
\end{gather}
Although it is not obvious from the expression \eqref{LOdeltaChiTerms}, the fact that $\delta \alpha^{\LO}$ and $\delta \phi^{\LO}$ satisfy the constraint \eqref{BoundaryDataConstraint} on the boundary data is essential to canceling some of the $r^2$ terms in the variation of the action. If we repeat this calculation with independent variations $\delta \alpha^{\LO}$ and $\delta \phi^{\LO}$, the $r^2$ terms in $\delta I$ cannot be canceled for any choice of the coefficients $\{c_i\}$ in \eqref{ActionWithCT}. This is examined in more detail in the next section.

With the coefficients of the surface terms fixed, the action is
\begin{align}\label{MinimalAction}
		I = & \,\, \frac{1}{2\kappa^{2}} \int_{\MM_r} \nts \dd^{4}x \, \sqrt{-g} \left( \RS - 2\,\Lambda - \frac{1}{4}\,F^{\mu\nu}\,F_{\mu\nu} - \frac{m^2}{2}\,A^{\mu} A_{\mu}\right) 
		+ \frac{1}{\kappa^2} \int_{M_r} \nts \dd^{3}x\,\sqrt{-h}\,K \\ \nonumber
		& \,\, + \frac{1}{\kappa^{2}} \int_{M_r} \nts \dd^{3}x\,\sqrt{-h}\, \left( - \frac{5}{2\,\ell} + \frac{1}{2\,\ell}\,A^{a} A_{a} - \frac{\ell}{4}\,R + \frac{\ell}{16}\,\FF^{ab} \FF_{ab}\right) ~.
\end{align}
In addition to having the required behavior under the different variations of the fields, this action is also finite as $r \to \infty$ for solutions of the equations of motion. The exact value of the action will depend on the particular solution, but the common asymptotics of the fields is enough to show that the on-shell action does not contain surface terms proportional to positive powers of $r$. Using the equations of motion and integration-by-parts, the on-shell action can be written
\begin{align}\label{OnShellAction}
		I\,\big|_\ms{\EE=0} = & \,\, \frac{1}{2\kappa^{2}} \int_{\MM_r} \nts \dd^{4}x \, \sqrt{-g} \left(2\,\Lambda + \frac{m^2}{2}\,A^{\mu} A_{\mu}\right) + \frac{1}{\kappa^2} \int_{M_r} \nts \dd^{3}x\,\sqrt{-h}\,\left( - \frac{1}{4}\,n_{\mu}\,F^{\mu\nu}\,A_{\nu} +  K \right) \\ \nonumber
		& \,\, + \frac{1}{\kappa^{2}} \int_{M_r} \nts \dd^{3}x\,\sqrt{-h}\, \left( - \frac{5}{2\,\ell} + \frac{1}{2\,\ell}\,A^{a} A_{a} - \frac{\ell}{4} \,R + \frac{\ell}{16}\,\FF^{ab} \FF_{ab} \right) ~.
\end{align}
The surface terms are evaluated in the asymptotic region $r \gg \ell$, making potential divergences easy to identify. The bulk term, on the other hand, depends on the behavior of the fields on the whole spacetime. Luckily, divergences associated with the $r \to \infty$ limit may be identified by replacing the integrand with its asymptotic behavior, performing the integral over $r$, and extracting terms proportional to positive powers of $r$\,\footnote{This procedure will not correctly extract divergences of the form $\log (r/\ell)$ . We do not expect to encounter such terms in AL theories with $z=2$, but in other theories a more careful accounting of the divergences coming from the bulk term is needed.}. This reveals potential divergences proportional to $r^4$ and $r^2$, but in both cases the bulk and surface contributions cancel or combine to form a total derivative on the boundary that can be discarded. Thus, the minimal action \eqref{MinimalAction} is free of $r \to \infty$ divergences on-shell.

Earlier in this section, we mentioned that other sets of surface terms can be used to construct a minimal action. 
This is because the scalar $A^a A_a$ approaches a constant at spatial infinity
\begin{gather}
	\lim_{r \to \infty} A^{a} A_{a} = -1 ~.
\end{gather}
As a result, the $A^a A_a$ term in \eqref{ActionWithCT} could be replaced with $\sqrt{- A^a A_a}$\,, as in \cite{Ross:2009ar}, or some other sufficiently well-behaved function. Likewise, the coefficients of the $R$ and $\FF^{ab} \FF_{ab}$ terms could be promoted to functions of $A^a A_a$. Then the surface term in \eqref{ActionWithCT} would be
\begin{gather}\label{GeneralSurfaceTerm}
	\frac{1}{\kappa^{2}}\,\int_{M_r} \nts \dd^{3}x \sqrt{-h} \left( f_\ms{0}(A^c A_c) + f_\ms{1}(A^c A_c)\,R 
		+ f_\ms{2}(A^c A_c)\, \FF^{ab} \FF_{ab} \raisebox{13pt}{} \right) ~,
\end{gather}
The calculations in this section place conditions on the on-shell values of the functions $f_{n}$ and their first derivatives, and any set of functions that meets those conditions provides a minimal action. Of course,  \eqref{GeneralSurfaceTerm} is by no means the most general set of surface terms that could be used. Terms like $R^{ab}\,A_{a}A_{b}$ and $F^{a}{}_{c}\,F^{bc}\,A_{a}A_{b}$, where two-derivative functions of the fields are contracted with two-index tensors constructed from $A_a$, could also be included\,\footnote{Terms that involve four or more derivatives of the fields are not relevant. Compared to the zero- and two-derivative terms, they carry additional factors of $r^{-2}$ which suppress their contributions to the action.}. Without additional conditions on $\delta I$, or some principle that restricts the functional form of the action, the surface terms are not uniquely specified.

\subsection{An Extended Action That Allows Independent Variations of $\alpha^{\LO}$ and $\phi^{\LO}$}
\label{sec:OtherAction}

The space of field configurations allowed by a well-defined variational principle must include generic fields that differ from solutions of the theory at NNLO. The minimal action derived in the last section enlarges this space, to include fields that fall off more slowly than NNLO. Roughly speaking, finiteness of $\delta I$ for Hamilton-Jacobi variations of the fields implies that $\delta I$ vanishes on-shell for field variations that fall off faster than $r^{2}$, as long as $\delta \alpha = \delta \phi$. The condition on $\delta \alpha$ and $\delta \phi$ is needed because certain cancellations in the last section relied on the variations respecting the constraint on the boundary data.

It is possible to extend this further, and construct actions such that $\delta I$ vanishes on-shell for \textit{independent} variations of the fields that fall off faster than $r^{2}$ at spatial infinity. Then the space of allowed field configurations includes all fields that obey AL boundary conditions, even if they do not admit an asymptotic expansion of the form \eqref{SigmaExpansion}-\eqref{PhiExpansion}. This is equivalent to requiring that the action have a finite on-shell response to leading order variations of the fields that do not satisfy the constraint on the boundary data. For lack of a better name, an action with this property will be referred to as `extended'. As we will see in the next section, a variational principle based on an extended action lets us explore two inequivalent definitions of a boundary stress tensor.

An example of an extended action can be constructed by adding another set of surface terms to \eqref{ActionWithCT} 
\begin{align}\label{Action3}
	I_r =  \frac{1}{2\kappa^{2}} \int_{\MM_r} \nts \textrm{d}^{4}x \, \sqrt{-g} \,\bigg( & \,\RS - 2\,\Lambda - \frac{1}{4}\,F^{\mu\nu}\,F_{\mu\nu} - \frac{m^2}{2}\,A^{\mu} A_{\mu}\,\bigg) 
	+ \frac{1}{\kappa^2} \int_{M_r} \nts \textrm{d}^{3}x\,\sqrt{-h}\,K \\ \nonumber
	+ \frac{1}{\kappa^{2}} \int_{M_r} \nts \textrm{d}^{3}x\,\sqrt{-h}\, \bigg(\,&  c_{0} + c_{1}\,A^{a} A_{a} 
	+ c_{2}\,R + c_{3}\,\FF^{ab} \FF_{ab} + c_{4} \, R^{ab} A_{a} A_{b} + c_{5}\,R\,A^{a} A_{a} \\ \nonumber
	&\,  + c_{6}\,\FF^{ab} \FF_{ab}\,A^{c} A_{c} + c_{7} \FF^{a}{}_{c} \,\FF^{bc} \, A_{a} A_{b} \,\bigg) ~.
\end{align}
The on-shell variation of the action takes the same general form \eqref{NewActionVariation}, with the replacement $p^{ab} \to p^{ab} + \Delta p^{ab}$ and $p^{a} \to p^{a} + \Delta p^{a}$. The shift in $p^{ab}$ and $p^{a}$ come from varying the new surface terms in \eqref{Action3}, which changes the previous results \eqref{CTmetvar} and \eqref{CTvectvar} by
\begin{align}
  \Delta p^{ab} = 
	 & \,\, c_4\,\bigg(\, 
			\frac{1}{2}\,h^{ab}\,R^{cd}\,A_c A_d - R^a{}_{c}\,A^{c} A^{b} 
			- R_{c}{}^{b}\,A^{a} A^{c} + \frac{1}{2}\,\nabla_c \nabla^{a} (A^c A^b ) 
			+ \frac{1}{2}\,\nabla_c \nabla^{b} (A^a A^c )  \\ \nonumber
		& \, \qquad - \frac{1}{2}\, \nabla^c \nabla_c (A^a A^b) - \frac{1}{2}\, h^{ab} \, \nabla_c \nabla_d (A^c A^d) 
				\,\bigg) \\ \nonumber	
	 & 		+ c_5 \, \bigg(\, 
			\frac{1}{2}\,h^{ab}\,R\,A^c A_c - R\,A^a A^b - R^{ab}\,A_c A^c + \nabla^a \nabla^b (A^c A_c) 
			- h^{ab}\,\nabla^d \nabla_d (A^c A_c) \, \bigg) \\ \nonumber
	 &  	+ c_6 \, \bigg(\, \frac{1}{2}\,h^{ab}\,\FF^{cd}\FF_{cd} \, A^e A_e - 2\,\FF^{a}{}_{c} \FF^{bc}\,A^d A_d 
			- \FF^{cd} \FF_{cd}\,A^a A^b \, \bigg) \\ \nonumber
	 &		+ c_7 \, \bigg(\, \frac{1}{2}\,h^{ab}\,\FF^{c}{}_{e} \FF^{de}\,A_{c} A_{d} 
			- \FF^{a}{}_{c} \FF^{dc}\,A_{d} A^{b} - \FF^{b}{}_{c} \FF^{dc}\,A^{a} A_{d}
			- \FF^{ac} \FF^{bd} \, A_{c} A_{d} \, \bigg) \\ 
  \Delta p^{a} = 
	 & \,\, 2\,c_4\, R^{ab}\,A_{b} + 2\,c_5 \, R\,A^{a} + c_6\,\bigg(\,2\,\FF^{cd} \FF_{cd}\,A^{a} 
			+ 4\,\FF^{ab}\,\nabla_{b}(A^c A_c) - 4\,A^c A_c \,\nabla_{b} \FF^{ba} \, \bigg) \\ \nonumber	
	 &  + c_7\,\bigg(\, 
			2\,\FF^{a}{}_{c} \FF^{bc}\,A_b + 2\,A^c A_b \, \nabla_{c} \FF^{ab} + 2\,\FF^{ab}\,A_b\,\nabla_c A^c
			+ 2\,\FF^{ab}\,A^c \,\nabla_c A_b \\ \nonumber
	 & \, \qquad \quad + 2\,A^a A_b \, \nabla_c \FF^{bc} + 2\,\FF^{bc}\,A_b\,\nabla_c A^a 
			+ 2\,\FF^{bc} \, A^a \, \nabla_c A_b \, \bigg)
\end{align}
Repeating the calculations of the previous section, with $\delta \alpha^{\LO}$ and $\delta \phi^{\LO}$ now treated as independent variations, determines the coefficients of the surface terms to be
\begin{gather}\label{GeneralCoefficients}
	c_0 = - \frac{5}{2\,\ell} \qquad c_1 = \frac{1}{2\,\ell} \qquad c_2 = - \frac{5\,\ell}{16} \qquad 
	c_3 = \frac{\ell}{8} \qquad c_{4} = \frac{\ell}{16} \qquad c_{5} = -\frac{\ell}{16} \qquad 
	c_{6} = \frac{\ell}{16} - \frac{1}{2}\,c_7 \vphantom{\bigg|}~.	
\end{gather}
All but one of the coefficients $\{c_i\}$ have been fixed, with $c_7$ left undetermined. However, the action turns out to be independent of $c_7$ when evaluated on-shell. This is because the last two surface terms in \eqref{Action3} are proportional to each other, up to terms that vanish as $r \to \infty$, when evaluated on a solution of the equations of motion 
\begin{gather}\label{L6L7}
	\sqrt{-h}\,\FF^{a}{}_{c} \,\FF^{bc}\,A_a A_b \, \big|_{\EE=0} = \frac{1}{2}\,\sqrt{-h}\,\FF^{ab} \FF_{ab}\,A^{c} A_{c}\,\big|_{\EE=0} ~.
\end{gather}
This results in a cancellation in the on-shell action when $c_6$ is given by \eqref{GeneralCoefficients}
\begin{gather}
	\sqrt{-h}\,\left( c_6\, \FF^{ab} \FF_{ab}\,A^{c} A_{c}
		+ c_7 \, \,\FF^{a}{}_{c} \,\FF^{bc}\,A_a A_b\right)\,\bigg|_{\EE=0}
	= 	\sqrt{-h}\, \frac{\ell}{16} \,\FF^{ab} \FF_{ab}\,A^{c} A_{c} \,\bigg|_{\EE=0} ~.
\end{gather}
In fact, the contributions that $\FF^{ab} \FF_{ab}\,A^{c} A_{c}$ and $\FF^{a}{}_{c} \,\FF^{bc}\,A_a A_b$ make to $\Delta p^{ab}$ and $\Delta p^{a}$ satisfy equalities similar to \eqref{L6L7}, which leads to on-shell cancellations in $\delta I$ that remove terms proportional to $c_7$. Therefore, the last surface term in \eqref{Action3} will not contribute to any of our calculations, and we are free to set $c_7 = 0$. The extended action is
\begin{align}\label{ExtendedAction}
		I =  & \,\,\frac{1}{2\kappa^{2}} \int_{\MM} \nts \dd^{4}x \, \sqrt{-g} \left( \RS - 2\,\Lambda - \frac{1}{4}\,F^{\mu\nu}\,F_{\mu\nu} - \frac{m^2}{2}\,A^{\mu} A_{\mu}\right) 
		+ \frac{1}{\kappa^2} \int_{\dM} \nts \dd^{3}x\,\sqrt{-h}\,K \\ \nonumber
		& + \frac{1}{\kappa^{2}} \int_{\dM} \nts \dd^{3}x\,\sqrt{-h} \, \bigg( - \frac{5}{2\,\ell} + \frac{1}{2\,\ell}\,A^{a} A_{a} - \frac{5\,\ell}{16}\,R + \frac{\ell}{8}\,\FF^{ab} \FF_{ab} + \frac{\ell}{16}\,R^{ab}\,A_{a} A_{b} \\ \nonumber
	    & \qquad \qquad \qquad \qquad \qquad - \frac{\ell}{16}\,R\,A^{a} A_{a} +\frac{\ell}{16}\,\FF^{ab} \FF_{ab}\,A^{c} A_{c} \,\bigg) ~.
\end{align}
Like the minimal action \eqref{MinimalAction}, this action is free of $r \to \infty$ divergences when evaluated on-shell.

\section{Boundary Stress Tensors and Conserved Charges}
\label{sec:BSTandCC}

Given a variational principle, we can construct a boundary stress tensor and use it to compute the conserved charges associated with the asymptotic symmetries of the theory. There are two notions of a boundary stress tensor that seem relevant for theories with AL boundary conditions. 

The standard construction due to Brown and York \cite{Brown:1992br} gives the boundary stress tensor as the response of the on-shell action to a variation of the boundary metric. This construction can be adapted to non-compact spacetimes by working on a compact region $\MM_r$, as we did in the previous section, and varying the functional $I_r$ with respect to the metric on the boundary $M_r$
\begin{gather} \label{BrownYork1}
	\tau^{ab} = \frac{2}{\sqrt{-h}}\,\frac{\delta I_r}{\delta h_{ab}}\,\bigg|_{\EE=0} ~.
\end{gather}	
Then one can define the charge $Q[\xi]$ associated with an asymptotic symmetry $\xi^{a}$ as the flux of the current $\tau_{ab}\,\xi^{b}$ across a cut $\CC$ of spatial infinity 
\begin{gather}\label{BYCharge}
	Q_\ts{BY}[\xi] = \int_{\CC} \dd^{2}x\,\sqrt{\sigma_\ms{\CC}}\,u^{a} \,\tau_{ab} \,\xi^{b}~,
\end{gather}
with $u^{a}$ the timelike unit vector normal to $\CC$. When the metric is the only field with support at spatial infinity, the charges \eqref{BYCharge} are conserved and generate the asymptotic symmetries of the theory. However, the authors of \cite{Hollands:2005ya} showed that this approach must be modified when there are non-vanishing tensor fields at spatial infinity in addition to the metric. For a theory with AL boundary conditions, their construction (reviewed below) instructs us to replace $\tau^{ab}$ in \eqref{BYCharge} with an improved boundary stress tensor
\begin{gather}\label{HIM1}
	T^{ab} = \tau^{ab} + \theta^{a}A^{b}~,
\end{gather}
where $\theta^{a}$ is given by
\begin{gather}
	\theta^{a} = \frac{1}{\sqrt{-h}}\,\frac{\delta I}{\delta A_{a}}\,\bigg|_{\EE=0} ~.
\end{gather}
As we will show, \eqref{HIM1} is the only definition of a boundary stress tensor that is consistent with the minimal action defined in section \ref{sec:MinimalAction}. Both constructions can be used with the extended action derived in section \ref{sec:OtherAction}, and there is an interesting relationship between the charges in that case.

\subsection{The Brown-York Definition Does Not Work For Minimal Actions}

For a variational principle based on the minimal actions defined in \ref{sec:MinimalAction}, the leading order variations of $h_{tt} = - \alpha^{2}$ and $A_{t}=\phi$ are constrained by \eqref{BoundaryDataConstraint}. Since $h_{tt}$ and $A_{t}$ cannot be varied independently, it is not clear how to apply Brown and York's definition of the boundary stress tensor for these actions. 

One might choose to ignore this issue and simply calculate \eqref{BrownYork1} for a minimal action, hoping for the best. This exercise makes the problem much more concrete. The resulting tensor, when used in \eqref{BYCharge}, gives charges that are not defined when we take the $r \to \infty$ limit. To see this in detail, consider the on-shell variation of the action
\begin{gather}\label{BoundaryVariation}
	\delta I\,\big|_{\EE=0} = \int_{M_r} \bns \dd^3{x}\,\sqrt{-h}\,\bigg( \,\frac{1}{2}\,\tau^{ab}\,\delta h_{ab} + \theta^{a} \, \delta A_{a} \,\bigg) ~.
\end{gather}
The coefficients of the field variations are given by
\begin{gather}
	\tau^{ab} = 2\,\big(\,\pi^{ab} + p^{ab} \,\big) \qquad \qquad \theta^{a} = \pi^{a} + p^{a}  
\end{gather}
where $p^{ab}$ and $p^a$ were defined in (\ref{CTmetvar}) and (\ref{CTvectvar}) respectively. Both $\tau^{ti}$ and $\theta^i$ vanish on-shell (at least to NNLO), and the asymptotic expansions for the remaining components are\,\footnote{Note that these expansions are for quantities with raised indices, while most of the other asymptotic expansions throughout the paper are for tensors with lower indices.}
\begin{align}
	\tau^{ij} = & \,\, \left( \frac{\ell}{r}\right)^{2} \tau_{\LO}^{ij} + \left( \frac{\ell}{r}\right)^{4} \tau_{\NLO}^{ij} + \left( \frac{\ell}{r}\right)^{6} \tau_{\NNLO}^{ij} + \ldots \\
	\tau^{tt} = & \,\, \left( \frac{\ell}{r}\right)^{4} \tau_{\LO}^{tt} + \left( \frac{\ell}{r}\right)^{6} \tau_{\NLO}^{tt} + \left( \frac{\ell}{r}\right)^{8} \tau_{\NNLO}^{tt} + \ldots \\
	\theta^{t} = & \,\, \left( \frac{\ell}{r}\right)^{2} \theta_{\LO}^{t} + \left( \frac{\ell}{r}\right)^{4} \theta_{\NLO}^{t} + \left( \frac{\ell}{r}\right)^{6} \theta_{\NNLO}^{t} + \ldots
\end{align}
The requirement that $\delta I$ vanishes on-shell for independent NNLO variations of the fields, as in \eqref{NNLOvariations}, means that the leading order terms in all three expansions vanish. The NLO terms in the expansions are relevant when we consider the response of the action to the Hamilton-Jacobi variations of the fields \eqref{HJVariations}. In general, $r^2$ divergences appear in $\delta I$ when these terms are non-zero. Since $\sigma_{ij}$ can be varied at leading order independently of the other fields, finiteness of $\delta I$ implies $\tau^{ij}_{\NLO} = 0$. But the constraint \eqref{BoundaryDataConstraint}, which forces $\delta \alpha = \delta \phi$, means that the NLO terms in $\tau^{tt}$ and $\theta^{t}$ do not need to vanish individually. Instead, they satisfy 
\begin{gather} \label{NLOCancel}
	\frac{1}{2}\,\tau^{tt}_{\NLO}\,(-4\,e^{4\chi}) + \theta^{t}_{\NLO}\,(2\,e^{2\chi}) = 0~.
\end{gather}  
One can check that, generically, $\tau^{tt}_{\NLO}$ and $\theta^{t}_{\NLO}$ are non-zero. 

The fact that $\tau^{tt}_{\NLO} \neq 0$ for the minimal action causes problems when we attempt to compute the conserved charge associated with the asymptotic symmetry $\xi^{a}\,\partial_{a} = \partial_{t}$ using \eqref{BYCharge}. With our boundary conditions it is natural to take $\CC$ to be one of the constant $t$ surfaces $\Sigma_{t}$, in which case 
\begin{align}
	Q[\partial_t] = & \,\, \int_{\Sigma_t} \nts \dd^{2}x \, \sqrt{\sigma} \,u^{t}\,\tau_{tt}\,\xi^{t} ~.
\end{align}
At large $r$, the factor of $\sqrt{\sigma}$ grows like $r^{2}$, the factor of $u^{t} = \alpha^{-1}$ falls off as $r^{-2}$, and the leading behavior of $\tau_{tt}$ is
\begin{gather}
	\tau_{tt} = h_{tt} h_{tt} \tau^{tt} = \left(\frac{r}{\ell}\right)^{2}\,e^{8\chi}\,\tau^{tt}_{\NLO} + \ldots ~.
\end{gather}
Combining these factors, we have
\begin{gather}\label{BadBYQ}
	Q[\partial_t] =  \left(\frac{r}{\ell}\right)^{2} \int_{\Sigma_t} \nts \dd^{2}x \, \sqrt{\sigma^{\LO}} \,e^{6\chi}\,\tau^{tt}_{\NLO} + \ldots ~,
\end{gather}
where `$\ldots$' denotes terms of order $r^{0}$. Thus, attempting to force the Brown-York definition of the boundary stress tensor on a minimal action leads to charges that are not defined in the $r \to \infty$ limit. One might wonder if the integral in \eqref{BadBYQ} vanishes, leaving only the finite sub-leading part, but this is not the case. For AL solutions the integrand is  
\begin{gather}
	e^{6\chi}\,\tau^{tt}_{\NLO} = e^{2\chi}\,\left(-\frac{\ell}{8}\,\RR^{\LO} + \frac{\ell}{4}\,\DD^{2}\chi - \frac{\ell}{2}\,\DD^i \chi \, \DD_i \chi \right) ~.
\end{gather}
Even if we restrict our attention to boundary conditions where $\chi$ is a constant, requiring that the integral of $\RR^{\LO}$ should vanish is contrary to our original goal of studying asymptotically Lifshitz spacetimes with curved spatial sections. 

That the energy has  contributions from both the asymptotic part of the metric and the asymptotic part of the Proca field has been noted previously \cite{Copsey:2010ya}. In asymptotically flat or AdS cases, there is enough gauge freedom to remove the additional Proca part, whereas for AL spacetimes failure to include the Proca part results in a loss of diffeomorphism invariance \cite{Copsey:2010ya}. In this sense it is not surprising that variations of metric and Proca cannot be considered independently. The Brown-York construction is simply the wrong approach for a minimal action.

\subsection{The Hollands-Ishibashi-Marolf Boundary Stress Tensor}

It is not surprising that the Brown-York approach fails for minimal actions, since there is a constraint that prevents $h_{tt}$ from being varied independently of $A_{t}$. Instead, we must employ the results of Hollands, Ishibashi, and Marolf (HIM), who showed in \cite{Hollands:2005ya} that the construction of conserved charges should be modified when other tensor fields besides the metric have support at spatial infinity.

To define the HIM boundary stress tensor, we first introduce a set of frame fields in the asymptotic region. In terms of these fields, the metric and vector are
\begin{gather}
	h_{ab} = \eta\indices{_A_B}\,e\indices{_a^A}\,e\indices{_b^B} \qquad A_{a} = \omega\indices{_A}\,e_{a}^{\,\,A}
\end{gather}
for a fixed metric $\eta\indices{_A_B}$, and a vector $\omega\indices{_A}$. The HIM boundary stress tensor is then defined as
\begin{gather}
	T\indices{^a_A} = \frac{1}{\sqrt{-h}}\,\frac{\delta I}{\delta e\indices{_a^A}}\,\bigg|_{\EE=0} ~.
\end{gather}
Note that there is no factor of `2', since we vary with respect to $e\indices{_a^A}$ instead of $h_{ab}$. The on-shell variation of the action now takes the form
\begin{gather}
	\delta I = \int_{M_r} \bns \dd^3{x}\,\sqrt{-h}\,\bigg( \, T\indices{^a_A} \,\delta e\indices{_a^A} + V^{A} \, \delta \omega_{A} \,\bigg) ~,
\end{gather}
with $T\indices{^a_A}$ and $V^{A}$ related to $\tau^{ab}$ and $\theta^{a}$ by
\begin{gather}
	T\indices{^a_A} = \tau^{ab}\,\eta\indices{_A_B}\,e\indices{_b^B} + \theta^{a}\,\omega\indices{_A} \qquad \qquad
	V^{A} = \theta^{a}\,e\indices{_a^A} ~.
\end{gather}
To compute conserved charges, we convert the frame index on $T\indices{^a_A}$ to a spacetime index and then apply the usual construction. This gives
\begin{gather}\label{HIMCharges}
	Q_\ts{HIM}[\xi] = \int_{\CC} \dd^2 x \, \sqrt{\sigma_\ms{\CC}}\,u^a \, T_{ab} \, \xi^{b} ~,
\end{gather}
where $T^{ab}$ with two spacetimes indices is
\begin{gather}\label{HIM2}
	T^{ab} = T\indices{^{a}_A}\,e\indices{^{b}_B}\,\eta\indices{^A^B} = \tau^{ab} + \theta^{a}\,A^{b} ~.
\end{gather}
This combination of $\tau^{ab}$ and $\theta^{a}$ neatly avoids the problems that we encountered trying to force the Brown-York construction on a minimal action. With AL boundary conditions, the HIM and Brown-York stress tensors differ only in their $t$-$t$ components	 
\begin{gather}
	T^{tt} = \tau^{tt} + \theta^{t}\,A^{t}~,
\end{gather}
which is the relevant component when computing the conserved charge associated with the asymptotic symmetry
generated by $\partial_t$. The NLO term in the asymptotic expansion of $T^{tt}$ vanishes by \eqref{NLOCancel}, so there is no obstruction to defining charges like there was in the Brown-York approach. Expressions for the components of $T^{ab}$ at NNLO are given in appendix \ref{sec:TermsInST}.

Finiteness of the charges \eqref{HIMCharges} follows from \eqref{NLOCancel}, so it is directly tied to finiteness of $\delta I$ for Hamilton-Jacobi variations of the fields. In fact, the HIM boundary stress tensor is just the coefficient (up to a constant factor) of $\delta \chi$ in the variation of the action. This is perhaps not too surprising, since the kinematics of the theory, which in this case includes the constraint \eqref{BoundaryDataConstraint}, plays an important role in the analysis of \cite{Hollands:2005ya}.

\subsection{Brown-York, Revisited}

Although the Brown-York definition of the boundary stress tensor does not work for minimal actions, it can be applied to the extended action derived in section \ref{sec:OtherAction}. In that case $\delta I$ is finite for independent variations of the fields at leading order, which requires $\tau^{tt}_{\NLO} = 0$ and $\theta^{t}_{\NLO} = 0$ instead of the condition \eqref{NLOCancel}. As a result, the charges \eqref{BYCharge} are finite as $r \to \infty$. Furthermore, the fairly restrictive assumptions about the asymptotic form of the fields insures that the charges \eqref{BYCharge} are conserved. Expressions for the NNLO terms in the components of the HIM and Brown-York stress tensors, for the action \eqref{ExtendedAction}, are given in appendix D.

It is interesting to compare the Brown-York and HIM definitions of the energy, since both constructions can be applied to actions like \eqref{ExtendedAction}. Using \eqref{HIM2} we have
\begin{gather}
	Q_\ts{HIM}[\partial_t] = Q_\ts{BY}[\partial_t] + \int_{\CC} \dd^2 x \, \sqrt{\sigma_\ms{\CC}}\,u^a A_{a} \, \theta_{b} \, \xi^{b}
\end{gather}
As before, we will take $\CC$ to be one of the spatial surfaces $\Sigma_t$, so that $u^{t} = \alpha^{-1}$. The constraint \eqref{BoundaryDataConstraint} implies $u^a A_a \to 1$ at spatial infinity, so the relation between the two notions of energy becomes
\begin{gather}
	Q_\ts{HIM}[\xi] = Q_\ts{BY}[\xi] + \Theta
\end{gather}
with $\Theta$ given by 
\begin{align}\label{Theta}
	\Theta = &\,\, \int_{\Sigma_t} \nts \dd^{2}x \, \sqrt{\sigma} \, \theta_{t} \\ \nonumber
		= & \,\, - \int_{\Sigma_t} \nts \dd^{2}x \, \sqrt{\sigma^{\LO}}\,e^{4\chi}\,\theta^{t}_{\NNLO} ~.
\end{align}
For systems with a thermodynamical interpretation, a better way to think of this is to define a `chemical potential' $\Psi = - u^{a} A_{a}$ for $\Theta$ \cite{Cheng:2009df}. Then for spacetimes with a horizon at $r=r_\ts{H}$, regularity of the fields requires $\Psi(r_\ts{H}) = 0$, and the difference in chemical potential between the horizon and spatial infinity is $\Delta \Psi = 0 - (-1) = 1$. The HIM and Brown-York charges are related by
\begin{gather}
	Q_\ts{HIM}[\xi] = Q_\ts{BY}[\xi] + \Theta \, \Delta \Psi ~.
\end{gather}
In other words, if we take the Brown-York charge to be the standard internal energy of the system -- which is a function of $\Theta$ -- then the HIM charge is the thermodynamic potential that depends instead on the chemical potential $\Psi$.

\section{The Lifshitz Topological Black Hole}
\label{sec:LTBH}

As an example, we can use the results of the last two sections to compute the action and energy for the topological black hole solutions of \cite{Mann:2009yx}. Expressed in the coordinate gauge \eqref{4to3plus1}, the metric and massive vector for these solutions takes the form
\begin{gather}\label{LTBHmetric}
	g_{\mu\nu} dx^\mu dx^\nu = \left(\frac{\ell}{r}\right)^{2} dr^2 - \left(\frac{r}{\ell}\right)^4 f_{-}(r)^2 f_{+}(r)^2 dt^2 + \left(\frac{r}{\ell}\right)^2 f_{-}(r)^2\,\ell^{2} d\Sigma_{k}^{\,\,2} \\ \label{LTBHvector}
	A_{\mu} dx^{\mu} = \left(\frac{r}{\ell}\right)^{2} f_{+}(r)^{2}\,dt ~,
\end{gather}
where the functions $f_{\pm}(r)$ are 
\begin{gather}
	f_{\pm}(r) = 1 \pm \frac{k}{8}\,\left(\frac{\ell}{r}\right)^{2}~.
\end{gather}
The constant $k$ controls the curvature of the spatial slices, and takes the values $\pm 1$ or $0$. In those three cases the two-dimensional line element $d\Sigma_{k}^{\,\,2}$ is given by
\begin{align}\label{ksurface}
\ell^{2} d \Sigma_{k}{}^{2} = \sigma_{ij}^{\LO} dx^i dx^j \qquad d \Sigma_{k}{}^{2} = & \,\, \begin{cases} d\theta^{2} + \sin^2 \theta\,d\phi^2, & k = 1 \\ 
d\theta^{2} + \theta^{2} \, d\phi^2, & k = 0  \\ 
d\theta^{2} + \sinh^2 \theta \, d\phi^2, & k = -1 ~, \end{cases} 
\end{align}
and the scalar curvature of $\sigma_{ij}^{\LO}$ is
\begin{gather}
	\RR^{\LO} = \frac{2\,k}{\,\ell^{2}}~.
\end{gather}
Thus, the $k=0$ solution is just the original Lifshitz spacetime \eqref{SimpleLifshitzMetric}-\eqref{SimpleLifshitzVector} written in terms of polar coordinates on $\Sigma_t$, while the $k=-1$ solution corresponds to a black hole with a horizon at $r_\ts{H} = \ell/\sqrt{8}$ and hyperbolic spatial slices. The $k=1$ solution is pathological due to a naked singularity at $r=0$, and is therefore only of interest as a potential asymptotic completion of some other solution that resolves the singularity. 

In terms of the parameterization of the fields introduced in section \ref{sec:BCandABS}, the topological black hole solutions corresponds to AL boundary conditions with $\chi = 0$. The NLO and NNLO terms in the asymptotic expansion of the fields, which can be taken directly from \eqref{LTBHmetric} and \eqref{LTBHvector}, are
\begin{align}
\sigma_{ij}^{\NLO} = - \frac{k}{4}\,\sigma_{ij}^{\LO} \qquad & \qquad
\sigma_{ij}^{\NNLO} = \frac{k^2}{64}\,\sigma_{ij}^{\LO} \\
\alpha^{\NLO} =  0 \qquad & \qquad
\alpha^{\NNLO} =  - \frac{k^2}{64} \\
\phi^{\NLO} = \frac{k}{4} \qquad & \qquad
\phi^{\NNLO} = \frac{k^2}{64}~.
\end{align}
These expressions correspond to the results \eqref{SimpleSigma}-\eqref{SimplePhi} obtained at the end of \ref{sec:BCandABS}, with the transverse-traceless spatial tensor at NNLO given by $Y_{ij} = 0$.

First we will consider the minimal action \eqref{MinimalAction} for this solution, and calculate the HIM boundary stress tensor and conserved charges. The NNLO term in $T^{tt}$ is
\begin{gather}
	T^{tt}_{\NNLO} = \frac{3\,\ell^{3}}{128}\,\left( \RR^{\LO}\right)^2 = \frac{3\,k^2}{32\,\ell} ~,
\end{gather}
so the energy obtained from the HIM definition of the conserved charges is
\begin{align}\label{LTBHEnergy1}
	Q_\ts{HIM}[\partial_t] = & \,\, \int_{\Sigma_t}\nts \dd^{2}x \sqrt{\sigma^{\LO}}\,T^{tt}_{\NNLO} \\ \nonumber
	= & \,\, \ell^{2}\,\text{Vol}_k(\Sigma_t)\,\frac{3\,k^2}{32\,\ell\,\kappa^{2}} ~,
\end{align}
where $\text{Vol}_k(\Sigma_t)$ is the (dimensionless) volume of the surface $\Sigma_t$ with metric \eqref{ksurface}. The energy and on-shell action are both zero for the $k=0$ solution, as expected, so let us focus on the case $k=-1$. This solution has a horizon at $r_\ts{H} = \ell/\sqrt{8}$, and regularity of the metric implies a periodicity $\beta = T^{-1} = 4 \pi \ell$ in Euclidean time. If we take the entropy $S$ to be one-quarter of the horizon area in Planck units, then we find
\begin{gather}
	T\,S = \ell^{2}\,\text{Vol}_{-1}(\Sigma_t)\,\frac{1}{4\,\ell\,\kappa^{2}} ~.
\end{gather} 
Evaluating the action \eqref{MinimalAction} on the Euclidean section of the solution gives
\begin{gather}\label{LTBHAction1}
	I_\ts{E} = - \ell^{2}\,\text{Vol}_{-1}(\Sigma_t)\,\frac{5}{32\,\ell\,\kappa^{2}}~,
\end{gather}
which satisfies the expected relation between the action, energy, temperature, and entropy
\begin{gather}\label{ActionEnergyEntropy}
	I_\ts{E} =  \beta\,\left(Q_\ts{HIM}[\partial_t] - T\,S \right)~.
\end{gather}
It is important to point out that both the energy \eqref{LTBHEnergy1} and the on-shell action \eqref{LTBHAction1} are sensitive to the choice of surface terms in the minimal action, but the result \eqref{ActionEnergyEntropy} is not. 
 
Now consider the extended action \eqref{ExtendedAction}, which contains additional surface terms compared to the minimal action. These terms shift the value of the energy obtained via the HIM construction, which is now
\begin{gather}\label{LTBHEnergy2}
	Q_\ts{HIM}[\partial_t] = \ell^{2}\,\text{Vol}_{k}(\Sigma_{t}) \, \frac{k^2}{32\,\ell\,\kappa^{2}} ~.
\end{gather}
However, the action evaluated on the Euclidean section is shifted by the same amount relative to \eqref{LTBHAction1}, so the relation \eqref{ActionEnergyEntropy} is still satisfied. For the $k=0$ solution the action vanishes, as before, while for the $k=-1$ it is 
\begin{gather}
	I_\ts{E} = - \ell^{2}\,\text{Vol}_{-1}(\Sigma_t)\,\frac{7}{32\,\ell\,\kappa^{2}}~.
\end{gather}
Of course, the action \eqref{ExtendedAction} also lets us calculate the Brown-York stress tensor, which is given by
\begin{gather}\label{}
	\tau^{tt}_{\NNLO} = \frac{5\,\ell^{3}}{128\,\kappa^{2}}\,\left(\RR^{\LO}\right)^{2} + \frac{\ell^{3}}{16\,\kappa^{2}}\,\DD_i \DD^{i} \RR^{\LO} - \frac{2}{\ell\,\kappa^{2}}\,\phi^{\NNLO} ~.
\end{gather}
It is interesting that, unlike the HIM boundary stress tensor, this depends on the part of the solution at NNLO that is not fixed by the kinematics of the theory. The charge associated with $\partial_t$ obtained from this stress tensor is
\begin{gather}
	Q_\ts{BY}[\partial_t] = \ell^{2}\,\text{Vol}_{k}[\Sigma_t]\,\frac{k^2}{8\,\ell}~.
\end{gather}
To relate this to the HIM energy, we must also evaluate the quantity $\Theta$ in \eqref{Theta}. The NNLO term in the asymptotic expansion of $\theta^{t}$ is
\begin{gather}
	\theta^{t}_{\NNLO} = \frac{\ell^{3}}{32\,\kappa^{2}}\,\left( \RR^{\LO} \right)^{2} + \frac{3\,\ell^{3}}{64\,\kappa^{2}}\,\DD_i \DD^i \RR^{\LO} - \frac{2}{\ell\,\kappa^{2}}\,\phi^{\NNLO}~,
\end{gather}
which also depends on the dynamical part of the solution. Then $\Theta$ is
\begin{gather}
	\Theta = -\int_{\Sigma_t}\nts \dd^{2}x \sqrt{\sigma}\,\theta^{t}_{\NNLO} = - \ell^{2}\, \text{Vol}_{k}(\Sigma_t) \, \frac{3\,k^{2}}{32\,\ell\,\kappa^{2}} ~.
\end{gather}
The chemical potential for $\Theta$ is $\Psi = - u^{a}\,A_{a}$, which takes the form
\begin{gather}
	\Psi = - \frac{1}{\alpha}\,\phi = - \frac{f_{+}(r)}{f_{-}(r)}
\end{gather} 
for the solution \eqref{LTBHmetric}-\eqref{LTBHvector}. When $k=-1$ this is regular on the Euclidean section, vanishing at $r_\ts{H}$ and approaching $-1$ as $r \to \infty$. Thus, the Brown-York and HIM charges satisfy
\begin{gather}\label{HIMvsBY}
	Q_\ts{HIM}[\partial_t] = Q_\ts{BY}[\partial_t] + \Theta\,\Delta \Psi ~,
\end{gather}
where $\Delta \Psi = \Psi(r_\ts{H}) - \Psi(\infty)$ is the difference in chemical potential between the horizon and spatial infinity.

\section{Discussion}
\label{sec:Discussion}

There are several ways in which our analysis can be improved. The most obvious extension is a treatment of asymptotically Lifshitz spacetimes that does not rely on the assumptions \eqref{MetricAssump}-\eqref{VectorAssump}. Relaxing these conditions is challenging, since the $3 \to 2+1$ split becomes much more complicated, but this is just a technical difficulty and we do not anticipate any real difficulties. A broader treatment of AL spacetimes would have a significant impact on many of our results. First, there will be a larger class of field variations consistent with more general boundary conditions. Requiring $\delta I$ to have the appropriate properties for the full class of field variations will give additional conditions on the surface terms, and this may partially or entirely resolve the ambiguities encountered in section \ref{sec:VarPrin}. A second consequence of more general boundary conditions is that the Brown-York charges, if they can be defined, will no longer be conserved. The Hollands-Ishibashi-Marolf construction \cite{Hollands:2005ya} should work perfectly well in that case, but the charges will not admit an interpretation along the lines of \eqref{HIMvsBY}.

Another relevant question is how our results extend to other values of the critical exponent $z$. The main difficulty here is that the character of the asymptotic expansions depends on $z$. For example, NNLO terms and NLO-squared terms enter the asymptotic expansions with different powers of $r$ when $1 \leq z < 2$, but with the same power of $r$ when $z=2$. This issue affects every aspect of the analysis in sections \ref{sec:ALS} and \ref{sec:VarPrin}, making the prospect of a general result for arbitrary values of $z$ seem unlikely. Instead, we expect to find actions that apply for distinct ranges of the critical exponent where functions of the fields have qualitatively similar asymptotic expansions. It would be interesting to see if there is a suitable action for some range of $z$ that includes $z=1$, since this corresponds to asymptotically AdS$_{4}$ spacetimes where the massive vector does not have support at spatial infinity.
The surface terms in the action are unique and well-understood in that case \cite{Balasubramanian:1999re, Emparan:1999pm}.

A further avenue of study would be to understand the relationship between our results and the stability of
AL spacetimes.  An initial value analysis recently showed that a generic normalizable state in an AL spacetime  will evolve in such a way to violate Lifshitz asymptotics in finite time \cite{Copsey:2010ya}. Whether or not the appropriate counterterm action can shed further light on this subject remains to be seen.

Finally, we have focused on the action as it relates to the gravitational theory, emphasizing the criteria for a well-defined variational problem and the definition and application of various boundary stress tensors. Despite the fact that we were originally motivated by a duality between gravitational and condensed matter theories, we have not discussed our results as they might apply to the CM dual. Once we have obtained a sufficiently general definition of AL boundary conditions, it will be interesting to see how the ambiguities in the action -- if they still remain -- relate to the structure of 1-, 2-, and higher n-point functions of operators in the dual CM theory.

~

\noindent{\bf Acknowledgments}

\noindent This work was supported in part by the Natural Sciences and Engineering Research Council of Canada, and by Loyola University Chicago's Summer Research Stipend Program. Many of the calculations in this paper were checked using \textit{xAct} \cite{Garcia:2011}, a tensor computer algebra package for \textit{Mathematica}.

~

\noindent {\bf Note:} As this paper was being prepared for submission, the papers \cite{Ross:2011gu, Baggio:2011cp} appeared on the arXiv. They take a systematic approach to holographic renormalization, along the lines of \cite{Papadimitriou:2004rz, Papadimitriou:2005ii}, for general values of $z$ and with a different definition of asymptotically Lifshitz boundary conditions than the ones we present in section \ref{sec:ALS}. Our results seem to agree with theirs where they overlap.

\appendix 

\section{The $4 \to 3+1$ Decomposition}
\label{sec:4to3plus1}
In section \ref{sec:Decompositions} we carry out a $3+1$ split of the fields and equations of motions, working in the coordinate gauge \eqref{4to3plus1}. Projections orthogonal to the surface $M_{r}$ are obtained by contracting with the unit normal vector $n^{\mu} = (\ell/r)\,\delta^{\mu}{}_{r}$, and projections tangent to $M_{r}$ with $P_{a}{}^{\mu} = \partial x^{\mu}/\partial x^{a}$. The metric (first fundamental form) on $M_{r}$ is given by the projection of the spacetime metric
\begin{gather}
	h_{ab} = P_{a}{}^{\mu} P_{b}{}^{\nu} g_{\mu\nu}~,
\end{gather}
and the extrinsic curvature (second fundamental form) is proportional to the Lie derivative of $h_{ab}$ along the normal
\begin{gather}
	K_{ab} = P_{a}{}^{\mu} P_{b}{}^{\nu} \,\CD_{\mu} n_{\nu} = \frac{1}{2}\,\Lie_{n} h_{ab} ~.
\end{gather}
The projection $P_{a}{}^{\mu}$ naturally defines a covariant derivative $\nabla_{a}$ that acts on tensors tangent to $M_{r}$ and is compatible with the metric $h_{ab}$. It is obtained from the complete projection (tangent to $M_{r}$) of $\CD_{\mu}$ acting on the tensor. For instance, given a vector $U_{\mu}$ such that $U_{\mu} n^{\mu} = 0$, the derivative is
\begin{gather}
	\nabla_{a} U_{b} \defeq P_{a}{}^{\mu}\,P_{b}{}^{\nu} \, \CD_{\mu} U_{\nu} ~.
\end{gather}
For an arbitrary 4-vector $V_{\mu}$ with components $V_{n} = n^{\mu} V_{\mu}$ and $V_{a} = P_{a}{}^{\mu} \,V_{\mu}$ we have
\begin{align}
	n^{\mu} n^{\nu} \, \CD_{\mu} V_{\nu} = & \,\, \Lie_{n} V_{n} \\
	P_{b}{}^{\nu} n^{\mu} \, \CD_{\mu} V_{\nu} = & \,\, \Lie_{n} V_{b} - V_{a}\,K^{a}{}_{b} \\
	P_{a}{}^{\mu} P_{b}{}^{\nu} \, \CD_{\mu} V_{\nu} = & \,\, \nabla_{a} V_{b} + V_{n}\,K_{ab} \\
	\CD_{\mu} V^{\mu} = & \,\, \Lie_{n} V_{n} + V_{n} \, K + \nabla_{a} V^{a}~.
\end{align}
Likewise, for the anti-symmetric tensor $F_{\mu\nu}$ with components given by \eqref{FieldStrengthPP}-\eqref{Ba}, the relevant projections are
\begin{align}
	n^{\nu}\,\CD^{\mu} F_{\mu\nu} = & \,\, - \nabla_{a} B^{a} \\
	P_{b}{}^{\nu} \,\CD^{\mu} F_{\mu\nu} = & \,\, \Lie_{n}B_{b} + B_{b}\,K - 2\,B_{a}\,K^{a}{}_{b} + \nabla^{a} \FF_{ab} ~.
\end{align}
These results are specific to the coordinate gauge \eqref{4to3plus1}. In a generic coordinate system they would acquire additional terms involving the vector $P_{a}{}^{\mu}\,n^{\nu}\,\CD_{\nu} n_{\mu}$ and its derivatives. 

To derive the $3+1$ equations of motion we also need the projections of the four-dimensional Ricci tensor normal and tangent to $M_r$. They can be written in terms of the extrinsic curvature $K_{ab}$ and the intrinsic Ricci tensor $R_{ab}$ as
\begin{align} \label{NNR}
n^{\mu} \,n^{\nu} \, \RS_{\mu\nu} = & \,\, - \Lie_{n}K - K^{ab}\,K_{ab} \\ \label{PNR}
P_{a}{}^{\nu} \, n^{\mu} \, \RS_{\mu\nu} = & \,\, \nabla^{b} K_{ab} - \nabla_{a} K \\ \label{PPR}
P_{a}{}^{\mu} P_{b}{}^{\nu} \, \RS_{\mu\nu} = & \,\, R_{ab} - \Lie_{n} K_{ab} - K\,K_{ab} + 2\,K_{a}{}^{c}\,K_{bc}~.
\end{align}
Using these projections to rewrite the trace of the Ricci tensor gives the four-dimensional Ricci scalar in terms of the intrinsic and extrinsic curvature
\begin{align}
\RS = & \,\, R - K^{2} - K^{ab}\,K_{ab} - 2\,\Lie_{n}K ~.
\end{align}

\section{The $3 \to 2+1$ Decomposition}
\label{sec:3to2plus1}

The $2+1$ split in section \ref{sec:Decompositions} involves projections orthogonal to $\Sigma_{t}$ with the (forward-pointing) timelike unit vector $u^{a} = \alpha^{-1}\,\delta^{a}{}_{t}$, and tangent to $\Sigma_t$ with $P_{i}{}^{a} = \partial x^{a}/\partial x^{i}$. The results are greatly simplified by the restrictions placed on the field in \eqref{MetricAssump}-\eqref{VectorAssump}. In particular, the extrinsic curvature of $\Sigma_{t} \subset M_r$ vanishes
\begin{gather}
	\theta_{ij} = - P_{i}{}^{a} P_{j}{}^{b}\,\nabla_{a} u_{b}
			    = - \frac{1}{2\,\alpha}\,\left(\partial_{t} \sigma_{ij} - D_i \beta_j - D_j \beta_i \right) = 0 ~,
\end{gather}
where $D_{i}$ is the two-dimensional covariant derivative that acts on tensors tangent to $\Sigma_t$. 
On the other hand, the acceleration vector $u^{b}\,\nabla_{b} u_{a}$ is non-zero if the lapse varies over $\Sigma_{t}$
\begin{gather}
	P_{i}{}^{a} \left( u^{b}\,\nabla_{b} u_{a} \right) = \frac{1}{\alpha}\,D_{i} \alpha ~.
\end{gather}
The projections of the three-dimensional Ricci tensor are 
\begin{align}
	u^{a} \, u^{b} \, R_{ab} = & \,\, \frac{1}{\alpha}\,D_i D^i \alpha \\
	P_{i}{}^{a} \, u^{b} \, R_{ab} = & \,\, 0 \\
	P_{i}{}^{a} \, P_{j}{}^{b} \, R_{ab} = & \,\, \RR_{ij} - \frac{1}{\alpha}\,D_{i} D_{j} \alpha  ~,
\end{align}
where $\RR_{ij}$ is the two-dimensional Ricci tensor for the metric $\sigma_{ij}$. The three-dimensional Ricci scalar is
\begin{gather}
	R = \RR - \frac{2}{\alpha}\,D_i D^i \alpha ~.
\end{gather} 
The remaining projections of fields and covariant derivatives are straight-forward.

\section{Transverse-Traceless Tensors Constructed from the Boundary Data}
\label{sec:TTBD}
The integral of the Ricci scalar is a topological invariant in two dimensions. As a result, the Einstein tensor vanishes identically in two dimensions
\begin{gather}\label{Bianchi2D}
  \RR_{ij} - \frac{1}{2}\,\sigma_{ij}\,\RR = 0 
\end{gather}
for any metric $\sigma_{ij}$. Now consider a metric $\sigma_{ij} + \varepsilon\,\gamma_{ij}$, where $\gamma_{ij}$ is some well-behaved but otherwise arbitrary tensor, and $\varepsilon$ is a small parameter. Expanding the Einstein tensor for this metric in powers of $\varepsilon$ gives
\begin{align}\label{ExpandEinstein}
	G_{ij}[\sigma + \varepsilon\,\gamma] = G_{ij}[\sigma] + \frac{\varepsilon}{2}\,\bigg( & \, D^{k} D_{i} \gamma_{kj} + D^{k} D_{j} \gamma_{ik} - D^k D_k \gamma_{ij} - D_i D_j \gamma^{k}{}_{k} - \RR\,\gamma_{ij} \\ \nonumber
	& \,\, + \frac{1}{2}\,\sigma_{ij}\,\RR\,\gamma^{k}{}_{k} - \sigma_{ij} \, D^{k} D^{l} \gamma_{kl} + \sigma_{ij} \, D^{k} D_{k} \gamma^{l}{}_{l} \, \bigg) + \OO(\varepsilon^{2})
\end{align}
where $D_{k}$ is the covariant derivative compatible with $\sigma_{ij}$, and indices are lowered and raised using $\sigma_{ij}$ and its inverse. The Ricci tensor vanishes identically in two-dimensions, so the left-hand side of \eqref{ExpandEinstein} and the first term on the right-hand side are both zero. And since $\varepsilon$ is a continuous parameter, the remaining terms in the Taylor expansion must vanish order-by-order.
For the $\OO(\varepsilon)$ term, this means
\begin{align}\label{LinearizedEinstein}
	0 = & \,\, D^{k} D_{i} \gamma_{kj} + D^{k} D_{j} \gamma_{ik} - D^k D_k \gamma_{ij} - D_i D_j \gamma^{k}{}_{k} - \RR\,\gamma_{ij} \\ \nonumber
	& \,\, + \frac{1}{2}\,\sigma_{ij}\,\RR\,\gamma^{k}{}_{k} - \sigma_{ij} \, D^{k} D^{l} \gamma_{kl} + \sigma_{ij} \, D^{k} D_{k} \gamma^{l}{}_{l} ~.
\end{align}
In other words, the linearized Einstein tensor vanishes for a small perturbation of the metric.

The result \eqref{LinearizedEinstein} is useful if we regard it as an identity that holds in two dimensions for a metric $\sigma_{ij}$ and an arbitrary symmetric tensor $\gamma_{ij}$. If we consider different tensors $\gamma_{ij}$ constructed from derivatives of a scalar function $\psi$, we obtain a set of identities that are useful when solving the equations of motion at NNLO in section \ref{sec:BCandABS}. First, let $\gamma_{ij}$ be
\begin{gather}
	\gamma_{ij} = D_i \psi D_j \psi ~.
\end{gather} 
Substituting this in \eqref{LinearizedEinstein}, commuting covariant derivatives, and applying \eqref{Bianchi2D}, we obtain
\begin{gather}\label{Identity1}
	0 = 2\,D_i D_j \psi \, D^k D_k \psi - 2\,D_i D^k \psi \, D_j D_k \psi + \sigma_{ij} \, D^k D^l \psi \, D_k D_l \psi - \sigma_{ij} \, D^k D_k \psi \, D^l D_l \psi ~.
\end{gather}
At first glance this combination of terms is not obviously zero, but the result can be confirmed by direct calculation. Another useful identity comes from setting 
\begin{gather}
	\gamma_{ij} = e^{c\,\psi} \, D_i \psi D_j \psi ~,
\end{gather} 
with $c$ a constant. Then \eqref{LinearizedEinstein}, supplemented with the identity \eqref{Identity1}, gives
\begin{align}\label{Identity2}
	0 = & \,\, D_i \psi \, D_j \psi \, D^k D_k \psi + D_i D_j \psi \, D^k \psi \, D_k \psi - D_i D_k \psi \, D_j \psi \, D^k \psi - D_j D_k \psi \, D_i \psi \, D^k \psi \\ \nonumber
	& \,\, - \sigma_{ij} \, D^k D_k \psi \, D^l \psi \, D_l \psi + \sigma_{ij}\,D^k D^l \psi \, D_k \psi \, D_l \psi ~. 
\end{align}
It is easy to check that the expressions on the right-hand sides of \eqref{Identity1} and \eqref{Identity2} are traceless. One can also check that they are transverse, which requires commuting covariant derivatives and multiple applications of \eqref{Bianchi2D}. 

There are two useful applications of the identities \eqref{Identity1} and \eqref{Identity2} in section \ref{sec:BCandABS}. First, solving the equations of motion for the NNLO term in the spatial metric is complicated, and applying identities like \eqref{Bianchi2D} at different stages in the calculation leads to expressions for $\sigma_{ij}^{\NNLO}$ which differ by linear combinations \eqref{Identity1} and \eqref{Identity2}. Second, these identities insure that the transverse-traceless tensor $Y_{ij}$ that appears at NNLO in the spatial metric does not depend on the boundary data. Consider a generic two-index symmetric tensor constructed from four-derivative combinations of $\sigma_{ij}$ and $\psi$, with at least one derivative acting on each factor of $\psi$. Thus, terms like $D_i \psi D_j \psi D^k D_k \psi$ or $\sigma_{ij}\,\RR\,D^k D_k \psi$ may appear, but not $\psi D_i D_j \psi D^k D_k \psi$ or $\sigma_{ij} \, \psi \, D^k D_k \RR$. The only transverse and traceless tensors of this sort are given by the terms on the right-hand sides of \eqref{Identity1} and \eqref{Identity2}, which vanish. Therefore, there are no transverse-traceless four-derivative functions of the boundary data that can appear in $\sigma_{ij}^{\NNLO}$.

\section{NNLO Terms in the Boundary Stress Tensors}
\label{sec:TermsInST}

We present here expressions for the components of the boundary stress-tensors for the
minimal action \eqref{MinimalAction} and the extended action \eqref{ExtendedAction}.

\noindent {\bf HIM boundary stress tensor}

The NNLO terms in the components of the HIM stress tensor for the minimal action \eqref{MinimalAction} are:
\begin{align}
	T^{tt}_{\NNLO} = & \,\, e^{-4\chi}\,\bigg( \,\frac{3\,\ell^{\,3}}{128}\,(\RR^{\LO})^2 
		- \frac{\,\ell^{\,3}}{32}\,\RR^{\LO}\,\DD^{k}\DD_{k}\chi 
		- \frac{\,\ell^{\,3}}{16}\,\DD_k \RR^{\LO} \, \DD^k \chi 
		- \frac{\,\ell^{\,3}}{8}\,\RR^{\LO}\,\DD^k \chi \DD_k \chi \\ \nonumber
	& 	+ \frac{3\,\ell^{\,3}}{32}\,\DD^k \DD_k \chi \, \DD^l \DD_l \chi 
		+ \frac{\,\ell^{\,3}}{8}\,\DD^k \DD_k \chi \, \DD^l \chi \,\DD_l \chi 
		- \frac{\,\ell^{\,3}}{8}\,\DD^k \chi \, \DD^l \DD_l \DD_k \chi \\ \nonumber
	& 	- \frac{\,\ell^{\,3}}{8}\,\DD^k \chi \DD_k \chi \, \DD^l \chi \DD_l \chi  
		- \frac{\,\ell^{\,3}}{2}\,\DD^k \DD^l \chi \, \DD_k \chi \, \DD_l \chi 
		- \frac{\,\ell^{\,3}}{4}\,\DD^k \DD^l \chi \, \DD_k \DD_l \chi \,\bigg)
\\
	T^{ij}_{\NNLO} = & \quad \frac{\,\ell^{\,3}}{128}\,\sigma^{ij}_{\LO}\,(\RR^{\LO})^2 
		- \frac{\,\ell^{\,3}}{16}\,\DD^i \DD^j \RR^{\LO} 
		+ \frac{\,\ell^{\,3}}{16}\,\sigma^{ij}_{\LO}\,\DD^k \DD_k \RR^{\LO} 
		- \frac{\,\ell^{\,3}}{8}\,\RR^{\LO}\,\DD^i \DD^j \chi \\ \nonumber
	&   + \frac{5\,\ell^{\,3}}{32}\,\sigma^{ij}_{\LO}\,\RR^{\LO}\,\DD^k \DD_k \chi 
		- \frac{\,\ell^{\,3}}{8}\,\left(\DD^i \chi \, \DD^j \RR^{\LO} + \DD^j \chi \, \DD^i \RR^{\LO} \right) 
		+ \frac{\,\ell^{\,3}}{4}\,\sigma^{ij}_{\LO}\,\DD^k \RR^{\LO}\,\DD_k \chi \\ \nonumber
	&   - \frac{\,\ell^{\,3}}{16}\,\sigma^{ij}_{\LO}\,\RR^{\LO}\,\DD^k \chi \, \DD_k \chi 
		- \frac{\,\ell^{\,3}}{8}\,\RR^{\LO}\,\DD^i \chi \, \DD^j \chi 
		- \frac{\,\ell^{\,3}}{8}\,\DD^i \DD^j \DD^k \DD_k \chi 
		+ \frac{\,\ell^{\,3}}{8}\,\sigma^{ij}_{\LO}\,\DD^k \DD_k \DD^l \DD_l \chi \\ \nonumber
	&   - \frac{\,\ell^{\,3}}{4}\,\left( \DD^i \chi \, \DD^j \DD^k \DD_k \chi 
					+ \DD^j \chi \, \DD^i \DD^k \DD_k \chi \right) 
		+ \frac{\,\ell^{\,3}}{2}\,\sigma^{ij}_{\LO}\,\DD_k \chi \, \DD^k \DD^l \DD_l \chi \\ \nonumber
	&   + \frac{3\,\ell^{\,3}}{4}\,\DD^i \DD^j \chi \, \DD^k \DD_k \chi 
		+ \frac{\,\ell^{\,3}}{2}\,\sigma^{ij}_{\LO}\,\DD^k \DD^l \chi \,\DD_k \DD_l \chi 
		- \frac{7\,\ell^{\,3}}{32}\,\sigma^{ij}_{\LO}\,\DD^k \DD_k \chi \, \DD^l \DD_l \chi \\ \nonumber
	&   + \frac{\,\ell^{\,3}}{4}\,\DD^i \chi \, \DD^j \chi \, \DD^k \DD_k \chi 
		+ \ell^{\,3} \, \DD^i \DD^j \chi \, \DD^k \chi \,\DD_k \chi 
		- \frac{\,\ell^{\,3}}{2}\,\left(\DD^i \chi \, \DD^j \DD_k \chi 
					+ \DD^j \chi \,\DD^i \DD_k \chi \right) \DD_k \chi \\ \nonumber
	&   + \frac{\,\ell^{\,3}}{2}\,\sigma^{ij}_{\LO}\,\DD^k \DD^l \chi \, \DD_k \chi \, \DD_l \chi 
		- \frac{5\,\ell^{\,3}}{8}\,\sigma^{ij}_{\LO}\,\DD^k \DD_k \chi \, \DD^l \chi \, \DD_l \chi
		+ \frac{\,\ell^{\,3}}{2}\,\DD^i \chi \, \DD^j \chi \, \DD^k \chi \, \DD_k \chi \\ \nonumber
	&   - \frac{3\,\ell^{\,3}}{8}\,\sigma^{ij}_{\LO}\,\DD^k \chi \, \DD_k \chi \, \DD^l \chi \DD_l \chi  
\end{align}
For the extended action \eqref{ExtendedAction}, the additional surface terms shift the NNLO terms in $T^{ab}$. In that case they are given by:
\begin{align}
	T^{tt}_{\NNLO} =  e^{-4\chi}\,
	\bigg( 
	& 	\,\,\frac{\,\ell^{\,3}}{128}\,(\RR^{\LO})^2 
		+ \frac{\,\ell^{\,3}}{64}\,\DD^k \DD_k \RR^{\LO}
		+ \frac{\,\ell^{\,3}}{16}\,\RR^{\LO}\,\DD^{k}\DD_{k}\chi 
		+ \frac{\,\ell^{\,3}}{16}\,\DD_k \RR^{\LO} \, \DD^k \chi 
		- \frac{\,\ell^{\,3}}{8}\,\RR^{\LO}\,\DD^k \chi \DD_k \chi \\ \nonumber
	& 	+ \frac{11\,\ell^{\,3}}{32}\,\DD^k \DD_k \chi \, \DD^l \DD_l \chi 
		+ \frac{3\,\ell^{\,3}}{8}\,\DD^k \DD_k \chi \, \DD^l \chi \,\DD_l \chi 
		+ \frac{\,\ell^{\,3}}{8}\,\DD^k \chi \, \DD^l \DD_l \DD_k \chi 
		+ \frac{\,\ell^{\,3}}{32}\,\DD^k \DD_k \DD^l \DD_l \chi \\ \nonumber
	& 	- \frac{\,\ell^{\,3}}{8}\,\DD^k \chi \DD_k \chi \, \DD^l \chi \DD_l \chi  
		- \frac{\,\ell^{\,3}}{2}\,\DD^k \DD^l \chi \, \DD_k \chi \, \DD_l \chi 
		- \frac{\,\ell^{\,3}}{4}\,\DD^k \DD^l \chi \, \DD_k \DD_l \chi 
	\,\bigg)
\end{align}
\begin{align} \label{TijExtended}
	T^{ij}_{\NNLO} = 
	& 	\quad \frac{\,\ell^{\,3}}{128}\,\sigma^{ij}_{\LO}\,(\RR^{\LO})^2 
		- \frac{\,\ell^{\,3}}{32}\,\DD^i \DD^j \RR^{\LO} 
		+ \frac{\,\ell^{\,3}}{32}\,\sigma^{ij}_{\LO}\,\DD^k \DD_k \RR^{\LO} 
		- \frac{\,\ell^{\,3}}{16}\,\RR^{\LO}\,\DD^i \DD^j \chi \\ \nonumber
	&   + \frac{3\,\ell^{\,3}}{32}\,\sigma^{ij}_{\LO}\,\RR^{\LO}\,\DD^k \DD_k \chi 
		- \frac{3\,\ell^{\,3}}{32}\,\left(\DD^i \chi \, \DD^j \RR^{\LO} 
					+ \DD^j \chi \, \DD^i \RR^{\LO} \right) 
		+ \frac{5\,\ell^{\,3}}{32}\,\sigma^{ij}_{\LO}\,\DD^k \RR^{\LO}\,\DD_k \chi \\ \nonumber
	&   + \frac{\,\ell^{\,3}}{32}\,\sigma^{ij}_{\LO}\,\RR^{\LO}\,\DD^k \chi \, \DD_k \chi 
		- \frac{\,\ell^{\,3}}{4}\,\RR^{\LO}\,\DD^i \chi \, \DD^j \chi 
		- \frac{\,\ell^{\,3}}{16}\,\DD^i \DD^j \DD^k \DD_k \chi 
		+ \frac{\,\ell^{\,3}}{16}\,\sigma^{ij}_{\LO}\,\DD^k \DD_k \DD^l \DD_l \chi \\ \nonumber
	&   - \frac{3\,\ell^{\,3}}{16}\,\left( \DD^i \chi \, \DD^j \DD^k \DD_k \chi 
					+ \DD^j \chi \, \DD^i \DD^k \DD_k \chi \right) 
		+ \frac{5\,\ell^{\,3}}{16}\,\sigma^{ij}_{\LO}\,\DD_k \chi \, \DD^l \DD_l \DD^k \chi \\ \nonumber
	&   - \frac{\,\ell^{\,3}}{8}\,\DD^i \DD^j \chi \, \DD^k \DD_k \chi 
		+ \frac{5\,\ell^{\,3}}{32}\,\sigma^{ij}_{\LO}\,\DD^k \DD_k \chi \, \DD^l \DD_l \chi \\ \nonumber
	&   + \ell^{\,3} \, \DD^i \DD^j \chi \, \DD^k \chi \,\DD_k \chi 
		- \frac{\,\ell^{\,3}}{2}\,\left(\DD^i \chi \, \DD^j \DD_k \chi 
					+ \DD^j \chi \,\DD^i \DD_k \chi \right) \DD_k \chi \\ \nonumber
	&   + \frac{\,\ell^{\,3}}{2}\,\sigma^{ij}_{\LO}\,\DD^k \DD^l \chi \, \DD_k \chi \, \DD_l \chi 
		- \frac{5\,\ell^{\,3}}{8}\,\sigma^{ij}_{\LO}\,\DD^k \DD_k \chi \, \DD^l \chi \, \DD_l \chi
		+ \frac{\,\ell^{\,3}}{2}\,\DD^i \chi \, \DD^j \chi \, \DD^k \chi \, \DD_k \chi \\ \nonumber
	&   - \frac{3\,\ell^{\,3}}{8}\,\sigma^{ij}_{\LO}\,\DD^k \chi \, \DD_k \chi \, \DD^l \chi \DD_l \chi  ~.
\end{align}

\noindent {\bf Brown-York boundary stress tensor}

The Brown-York stress tensor can also be defined for the extended action \eqref{ExtendedAction}. In that case the component $\tau^{ij}_{\NNLO}$ is the same as \eqref{TijExtended}, and the $t$-$t$ component is given by
\begin{align}
	\tau^{tt}_{\NNLO} = e^{-4\chi}\,
	\bigg( 
	& 	\,\,\frac{5\,\ell^{\,3}}{128}\,(\RR^{\LO})^2 
		- \frac{2}{\ell}\,\phi^{\NNLO}
		+ \frac{\,\ell^{\,3}}{16}\,\DD^k \DD_k \RR^{\LO}
		+ \frac{\,\ell^{\,3}}{16}\,\RR^{\LO}\,\DD^{k}\DD_{k}\chi 
		+ \frac{3\,\ell^{\,3}}{16}\,\DD_k \RR^{\LO} \, \DD^k \chi \\ \nonumber
	&	+ \frac{\,\ell^{\,3}}{8}\,\RR^{\LO}\,\DD^k \chi \DD_k \chi 
		+ \frac{11\,\ell^{\,3}}{32}\,\DD^k \DD_k \chi \, \DD^l \DD_l \chi 
		+ \frac{9\,\ell^{\,3}}{8}\,\DD^k \DD_k \chi \, \DD^l \chi \,\DD_l \chi 
		+ \frac{3\,\ell^{\,3}}{8}\,\DD^k \chi \, \DD^l \DD_l \DD_k \chi \\ \nonumber
	& 	+ \frac{\,\ell^{\,3}}{8}\,\DD^k \DD_k \DD^l \DD_l \chi
		- \frac{\,\ell^{\,3}}{8}\,\DD^k \chi \DD_k \chi \, \DD^l \chi \DD_l \chi  
		- \frac{3\,\ell^{\,3}}{4}\,\DD^k \DD^l \chi \, \DD_k \chi \, \DD_l \chi 
		- \frac{3\,\ell^{\,3}}{8}\,\DD^k \DD^l \chi \, \DD_k \DD_l \chi 
	\,\bigg) ~.
\end{align}
We also require the NNLO term in the expansion of $\theta^{a}$. With AL boundary conditions the component $\theta^{i}$ vanishes, and $\theta^{t}_{\NNLO}$ is
\begin{align}
	\theta^{\,t}_{\NNLO} = e^{-2\chi}\,
	\bigg( 
	&	\,\, \frac{\,\ell^{\,3}}{32}\,(\RR^{\LO})^2
		- \frac{2}{\ell}\,\phi^{\NNLO}
		+ \frac{3\,\ell^{\,3}}{64}\,\DD^k \DD_k \RR^{\LO}
		+ \frac{\,\ell^{\,3}}{8}\,\DD_k \chi \, \DD^k \RR^{\LO}
		+ \frac{\,\ell^{\,3}}{4}\,\RR^{\LO}\,\DD^k \chi \,\DD_k \chi \\ \nonumber
	&	+ \frac{3\,\ell^{\,3}}{4}\,\DD^k \DD_k \chi \, \DD^l \chi \, \DD_l \chi
		+ \frac{\,\ell^{\,3}}{4}\,\DD^k \chi \, \DD^l \DD_l \DD_k \chi
		+ \frac{3\,\ell^{\,3}}{32}\,\DD^k \DD_k \DD^l \DD_l \chi
		- \frac{\,\ell^{\,3}}{4}\,\DD^k \DD^l \chi \, \DD_k \chi \,\DD_l \chi \\ \nonumber
	&	- \frac{\,\ell^{\,3}}{8}\,\DD^k \DD^l \chi \, \DD_k \DD_l \chi
	\,\bigg)~.
\end{align}

\end{document}